\documentclass[aps,pre,reprint,superscriptaddress,10pt]{revtex4-2}
\usepackage{amsmath, amssymb, graphicx, color, braket,mathtools}
\usepackage{bbold, bm}
\usepackage{verbatim}

\begin{document}

\title{{Hidden symmetry in interacting-quantum-dot-based multi-terminal Josephson junctions}}

\author{Peter Zalom}
\email{zalomp@fzu.cz}
\affiliation{Institute of Physics, Czech Academy of Sciences, Na Slovance 2, CZ-18200 Praha 8, Czech Republic}

\author{M. \v{Z}onda}
\email{martin.zonda@matfyz.cuni.cz}

\affiliation{Department of Condensed Matter Physics, Faculty of Mathematics and Physics, Charles University, Ke Karlovu 5, CZ-121 16 Praha 2, Czech Republic}

\author{T. Novotn\'y}
\email{tomas.novotny@matfyz.cuni.cz}
\affiliation{Department of Condensed Matter Physics, Faculty of Mathematics and Physics, Charles University, Ke Karlovu 5, CZ-121 16 Praha 2, Czech Republic}

\date{\today}

\begin{abstract}
We study a multi-terminal Josephson junction based on an interacting quantum dot coupled to $n$ superconducting BCS leads. Using an Anderson type model of a local level with an arbitrary onsite Coulomb repulsion, we uncover its surprising equivalence with an effective two-terminal junction with symmetric couplings to appropriately phase-biased leads. Regardless of the strength of the Coulomb interaction, this hidden symmetry enables us to apply well-established numerical and theoretical tools for exact evaluation of various physical quantities, and imposes strict relations among them. Focusing on three-terminal devices, we then demonstrate several phenomena such as the existence of the finite energy band crossings, superconducting transistor and diode effects, as well as current phase relation modulation. 
\end{abstract}


\maketitle

\textit{Introduction.}
Josephson junctions (JJs) serve as fundamental components for a range of quantum devices thanks to their precise superconducting phase control~\cite{Likharev-1991,Cleuziou-2006,Clarke-2008,Rodero-2011,Meden-2019}. Therefore, their multiterminal counterparts with $n \geq 3$ leads have recently received significant theoretical attention. From a topological perspective, their subgap energy levels span a synthetic $(n-1)$-dimensional Brillouin zone (BZ) leading to the potential emergence of zero- and finite-energy Weyl nodes~\cite{Heck-2014,Riwar-2016,Eriksson-2017,Lutchyn-2018,Xie-2018,Xie-2019,Repin-2022,Gavensky-2023}. Additionally, multi-terminal JJs with integrated topological superconductors hold promise for performing braiding operations on zero-energy Majorana bound states~\cite{Alicea-2011,Zazunov-2017}, while in the non-equilibrium regime, intriguing phenomena such as Cooper pair quartet transport emerge \cite{Freyn-2011}. On the experimental front, the realization of multi-terminal superconducting (SC) devices has recently advanced significantly, with pioneering experiments in graphene \cite{Draelos-2019}, weak links~\cite{Pankratova-2020} and ongoing innovations appearing~\cite{Graziano-2022,Gupta-2023,Melin-2023,Zhang-2023,Coraiola-2023,Sadashige-2023}.

Because of the computational constraints arising from the number of leads, interactions within the central junction region are mostly neglected or approximated~\cite{Meng-2009,Grove-Rasmussen-2018,Zonda-2023}. Notably, even the Numerical Renormalization Group (NRG), a standard for analyzing strongly interacting superconducting single-level Anderson impurity models (SC-AIM), whose relevance to accurately describe realistic experimental setups was established over a decade ago \cite{Pillet-2010,Luitz-2012}, faces challenges in this context \cite{Zalom-2023}.

In this work, we show that the paradigmatic $n$-terminal SC-AIM can be {\em exactly} mapped onto a two-terminal version with symmetric tunnel couplings and a suitable phase bias. The mapping is completely determined by the original configuration through a gauge-invariant geometric factor denoted as $\chi$. Since the two-terminal SC-AIM can be exactly addressed by means of NRG \cite{Satori-1992, Sakai-1993, Hecht-2008, Zalom-2023}
and Quantum Monte Carlo \cite{Siano-2004,Siano-2005rep, Luitz-2012,Gubernatis-2016}, or by other numerous tools~\cite{Karrasch-2008,Meng-2009,Grove-Rasmussen-2018,Zonda-2022,Zonda-2023,Paaske-arxiv}, the knowledge of the geometric factor $\chi$ can be used to completely understand the behavior of multi-terminal SC-AIM as demonstrated here. 

Unveiling the geometric properties of the solution for multi-terminal SC-AIM allows us to make meaningful and nontrivial assertions about its phase diagrams and associated Josephson currents. We especially highlight the practical possibility to realize the high-symmetry points, which are related to the so-called doublet chimney~\cite{Bargerbos-2022,Zitko-arxiv,Paaske-arxiv} and remain robust regardless of the strength of the Coulomb interaction. Furthermore, our work showcases the practical advantages of incorporating three-terminal quantum dot-based devices into SC circuits, as they introduce SC transistor and diode effects, and enable the modulation of supercurrent phases \footnote{The mechanism behind both effects is novel and relies on the full phase control over the three terminal architecture of the device and the presence of Coulomb interactions in the junction region.}.

\begin{figure}[ht]
	\includegraphics[width=0.85\columnwidth]{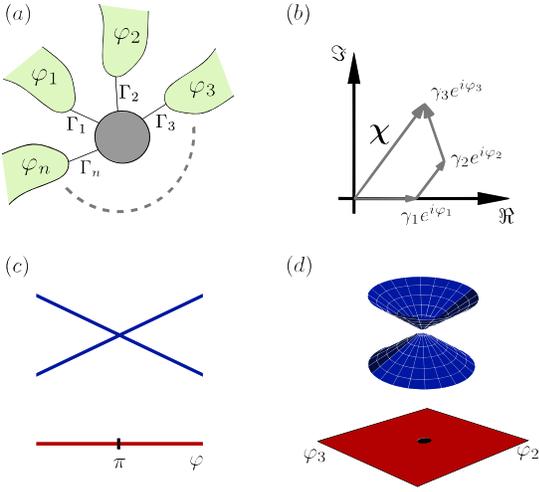}
	\caption{
		(a)
		Schematic depiction of a multi-terminal Josephson junction with $n$ SC leads (green) and a centrally-placed interacting region of a single level QD (gray).
		(b)
		The complex number $\bm{\chi}$ according to Eq.~\eqref{eq:chi}, 
		(shown for $n=3$) plays a fundamental role for properties of $n$-terminal SC-AIMs. When $\bm{\chi}=0$ finite energy crossings of two singlet states (blue) above a doublet ground state (red) are observed.
		(c) 
		Singlet crossing for $n=2$ occurs only under the highly restrictive condition $\Gamma_1=\Gamma_2$ at $\varphi \equiv \varphi_2-\varphi_1=\pi$ 
		(d) For $n=3$, the singlet crossings appear for $\Gamma_1,\,\Gamma_2,\,\Gamma_3$ satisfying a triangular inequality $\gamma_{\text{max}}\leq 1/2$  where $\gamma_{\text{max}}=\text{max}(\{\gamma_j\})$ and $\gamma_j\equiv \Gamma_j/\sum_l \Gamma_l$.
		\label{fig:model}}
\end{figure} 

\textit{Model:}
We consider a general multi-terminal SC-AIM with a single-level quantum dot (QD) as depicted schematically in Fig.~\ref{fig:model}(a). Its Hamiltonian reads
\begin{subequations}
\begin{align}
	H&=H_d+\sum_{j=1}^{n}\left(H_{j,\text{SC}}+H_{j,T}\right)
\end{align}
with  $j \in \{1,\ldots n \}$ denoting a given lead with an SC phase $\varphi_j$ and
\begin{align}
	H_d
	&=
	\sum_{\sigma} 
	\varepsilon_{d}
	d^{\dagger}_{\sigma}
	d^{\vphantom{\dagger}}_{ \sigma}
	+
	U
	d^{\dagger}_{\uparrow}
	d^{\vphantom{\dagger}}_{ \uparrow}
	d^{\dagger}_{\downarrow}
	d^{\vphantom{\dagger}}_{ \downarrow},
	\label{eq:dotH}
	\\
	H_{j,\text{SC}}
	&=
	\sum_{\mathbf{k}\sigma} 
	\, \varepsilon_{\mathbf{k}j}
	c^{\dagger}_{\mathbf{k} j\sigma}
	c^{\vphantom{\dagger}}_{\mathbf{k} j\sigma}
	-\sum_{\mathbf{k}}
	\left(
	\Delta_{j}
	c^{\dagger}_{\mathbf{k}j \uparrow} 
	c^{\dagger}_{-\mathbf{k}j \downarrow}
	+
	\textit{H.c.}\right),
	\label{eq:bcsH}
	\\
	H_{j,T} 
	&=
	\sum_{\mathbf{k} \sigma} \,
	\left(V^*_{\mathbf{k}j}  
	c^{\dagger}_{\mathbf{k}j\sigma}
	d^{\vphantom{\dagger}}_{\sigma} 
	+ 
	V_{\mathbf{k}j}  
	d^{\dagger}_{\sigma}
	c^{\vphantom{\dagger}}_{\mathbf{k}j\sigma}\right),
	\label{eq:tunnelH}
\end{align}
\end{subequations}
where $c^{\dagger}_{\mathbf{k} j\sigma}$ ($c^{\vphantom{\dagger}}_{\mathbf{k}j\sigma}$) creates (annihilates) an electron of spin $\sigma \in \{\uparrow, \downarrow \}$,  quasi-momentum $\mathbf{k}$, and energy $\varepsilon_{\mathbf{k}j}$ in the lead $j$, while $d^{\dagger}_{\sigma}$ ($d^{\vphantom{\dagger}}_{\sigma}$) creates (annihilates) a dot electron of spin $\sigma$. We assume the same gap $\Delta$ and the same band width $2B$ in all leads, so that $\Delta_{j} \equiv \Delta e^{i\varphi_{j}}$. The coupling to the leads is conveniently characterized by the tunneling strengths ($\hbar=1$) $\Gamma_j\equiv\pi\sum_{\mathbf{k}}|	V_{\mathbf{k}j}|^2\delta(\omega-\varepsilon_{\mathbf{k}j})$, which are presumed to be energy independent for simplicity. The total tunneling strength is $\Gamma\equiv\sum_{j=1}^{n}\Gamma_j$, while relative couplings are $\gamma_j\equiv\Gamma_j/\Gamma$ ( $\sum_{j=1}^{n}\gamma_j=1$). The QD is characterized by its energy level $\varepsilon_d$ and Coulomb repulsion $U$. In this letter, we focus on a half-filled QD by setting $\varepsilon_d=-U/2$, but stress that all of the findings can be easily extended beyond such a constraint.

\textit{Gauge invariance of the solution:}
When focusing only on physical quantities related to the dot degrees of freedom, like the on-dot spectral function or thermodynamic quantities such as the free energy and supercurrents, only the dot Green function is required. It is a functional of $U$ and the non-interacting Green function, which, in turn, is a functional of the (retarded) tunneling self-energy $\mathbb{\Sigma}(\omega^+)$ given by the $n$ leads. Using Nambu spinors $D^{\dagger} =  \left(d^{\dagger}_{\uparrow}, 
d^{\vphantom{\dagger}}_{\downarrow} \right)$, the matrix form of $\mathbb{\Sigma}(\omega^+)$ becomes
\begin{equation}
	\mathbb{\Sigma}(\omega^+)=	
	\Gamma 
	\begin{pmatrix} 	
		\omega & \bm{\chi} \Delta 
		\\ 
		\bm{\chi}^* \Delta	& \omega 
	\end{pmatrix}
	F(\omega^+),
	\label{eq:bcs_sigma_d_gauge}
\end{equation}
where $F(\omega^+)$ parametrically depends only on $\Delta$ and $B$ (see e.g., Ref.~\cite{Zalom-2021} and the Supplemental Material (SM) \cite{[{See Supplemental Material at }][{ for details on high symmetry-points and additional data on four terminal devices, which includes Refs.~\cite{Zalom-2021,Bauer-2007,Zalom-2021r,Zalom-2022,Meden-2019,Saldana-2018,NRGzenodo}.}]supp}). Complex-valued $\bm{\chi}$ reads
\begin{equation}\label{eq:chi}
\bm{\chi}   
\equiv     
\sum_{j=1}^{n} 
\gamma_j e^{i\varphi_j}
\end{equation}
and contains complete information about the geometric configuration of the $n$-terminal SC-AIM including all relative weights and phase biases. Using the global gauge invariance, we can moreover shift all phases as $\varphi_j \rightarrow \varphi_j-\delta$, which rotates $\bm{\chi}$ clock-wise by $\delta$, but leaves all physical properties invariant \footnote{The dot Green function actually changes, but it has no physical consequences, for a detailed account of this issue for the two-terminal setup see Ref.~\cite{Kadlecova-2017}, the situation for multi-terminal setup is completely analogous.}. Consequently, only the gauge-invariant magnitude $\chi \equiv |\bm{\chi}|$ is of significance, so the replacement $\bm{\chi} \rightarrow \chi$ can be readily performed in Eq.~\eqref{eq:bcs_sigma_d_gauge}. As shown in SM \cite{supp}, $\chi$ can be simplified to
\begin{equation}
	\chi
	\equiv
	|\bm{\chi}|	
	=
	\sqrt{1-4\sum_{j>l=1}^n	\gamma_j\gamma_l \sin^2 \frac{\varphi_j-\varphi_l}{2}}.
	\label{eq:chi_sqrt}
\end{equation}  
While generally $0 \leq \chi \leq 1$, for $n=1$ it trivially reads $\chi=1$. However, already a coupling-symmetric two-terminal SC-AIM encompasses all possible values of $\chi$ as $\chi=|\cos(\varphi/2)|$ \cite{Kadlecova-2017}. Consequently, {\em any} $n$-terminal setup can be mapped onto its coupling-symmetric two-lead counterpart with the same $U$, total $\Gamma$ and a phase difference $\varphi$ that corresponds to the multi-terminal value of $\chi$. This constitutes the main finding of our work with a number of conceptual and practical implications, which we will now explore.

\begin{figure*}[ht]
	\includegraphics[width=2.0\columnwidth]{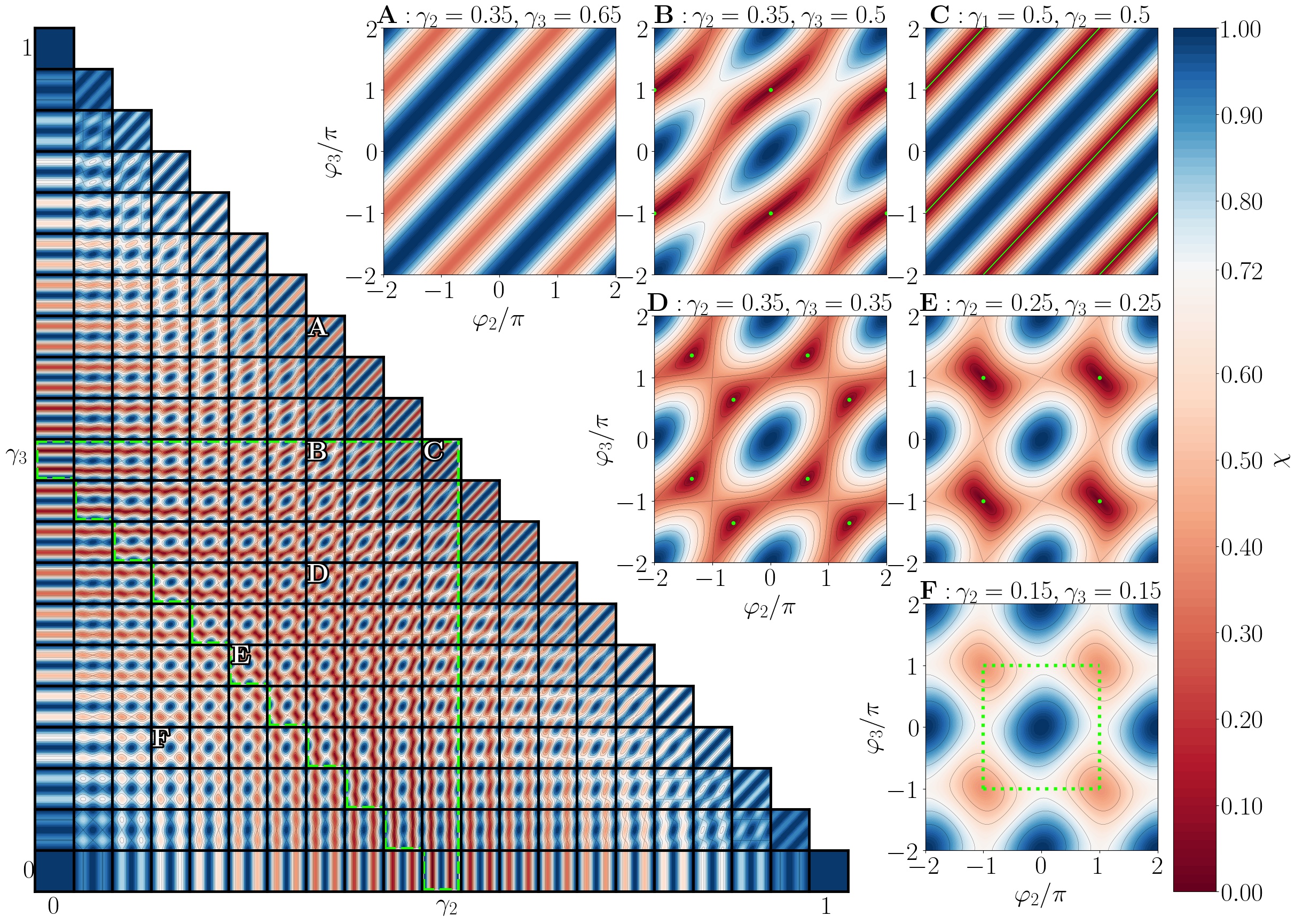}
	\caption{Maps of geometric factor $\chi$ (Eq.~\eqref{eq:chi_sqrt}) in $\varphi_2$ - $\varphi_3$ planes for a three-terminal setup ($\varphi_1 \equiv 0$) and different values of $\gamma_2$ and $\gamma_3$. The coloring is chosen so that the maps can be read as phase diagrams for the case $\Gamma=\Delta$, $U=3\Delta$ for which $\chi^*=0.721$. Here white represents the phase boundaries $\chi=\chi^*$, blue marks the singlet GS ($\chi>\chi^*$), and red a doublet GS ($\chi<\chi^*$). In the composite map on the left, the phase diagrams are ordered along the $\gamma_2$ and $\gamma_3$ axes with a step in $\gamma_2$ and $\gamma_3$ being $0.05$. The maps within the dashed green border satisfy the triangular rule $\text{max}(\{\gamma_j\})\leq1/2$ and can therefore host the high-symmetry points, where $\chi=0$. Four of such cases, e.g. {\bf B - E}, are shown enlarged with $\chi=0$ points indicated by green dots ({\bf B, D, E}) or solid lines ({\bf C}). Cases \textbf{A} and \textbf{F} are taken from the outside of the triangular region. Note that, for clarity, all maps extend beyond the first Brillouin zone, i.e.~ $\varphi_2,\varphi_3 \in ( -\pi,\pi \rangle$, which is marked by a dotted green square in panel {\bf F}.
	\label{fig:triangle}}
\end{figure*}

\begin{figure}[ht]
	\includegraphics[width=1.0\columnwidth]{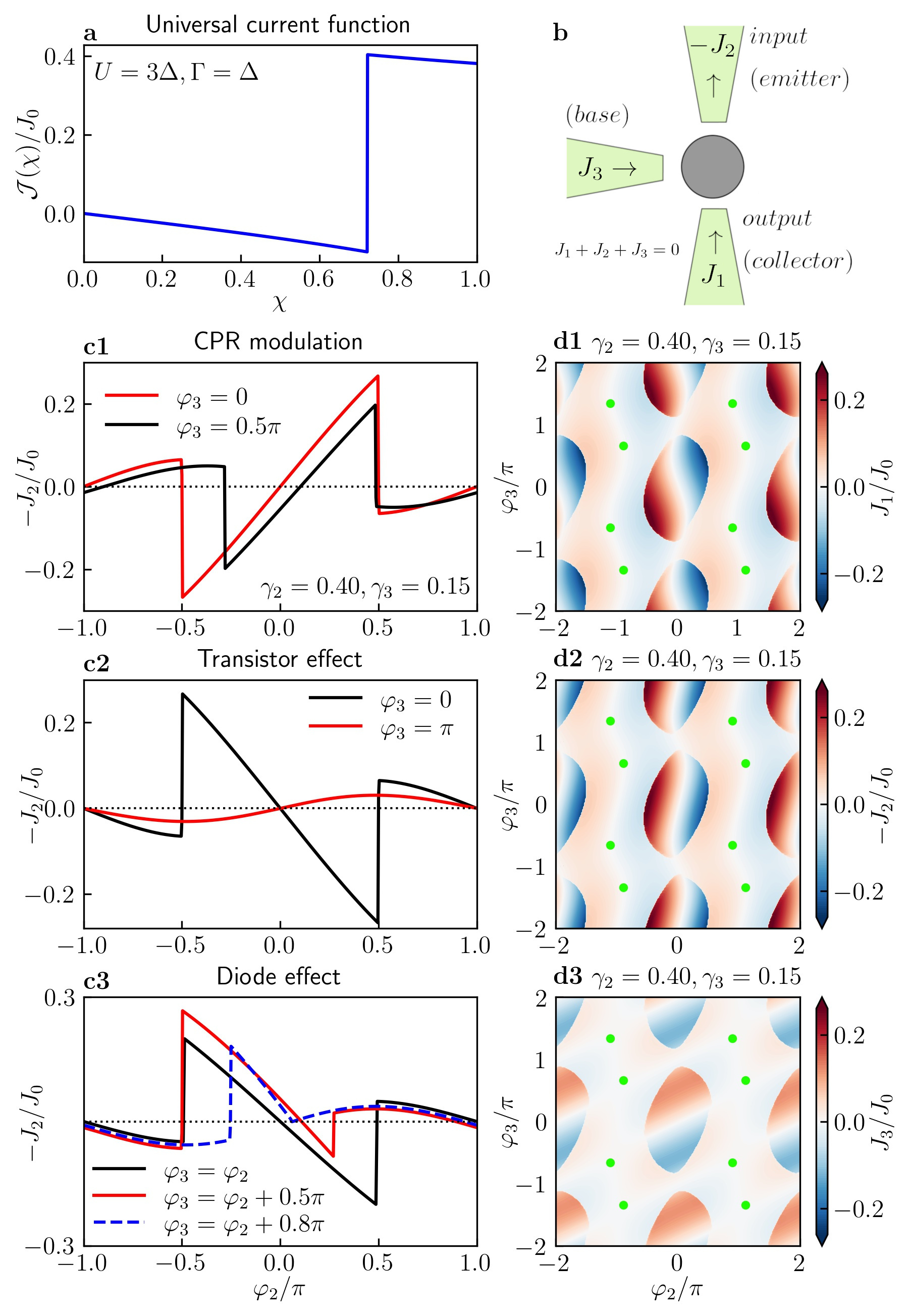}
	\caption{ 
	(a) 
	The universal function $\mathcal{J}(\chi)$ calculated for $U=3\Delta$ and $\Gamma=\Delta$ and scaled by $J_0 = 2\Delta e/\hbar$. 
	(b) 
	Scheme of the supercurrents $J_j$ in a three terminal device. 
	(c1) 
	The current phase relation (CPR) of the supercurent $J_2$ is modulated by controlling $\varphi_3$.
	(c2) 
	Supercurrent $J_2$ can be switched-off by phase-manipulation of the weakly coupled lead $j=3$ which demonstrates a superconducting transistor-like effect. 
	(c3)  
	$J_2$ as a function of $\varphi_2$ for $\varphi_3=\varphi_2 + \phi$ illustrates the appearance of a diode effect. 
	Maps (d1), (d2), (d3) 
	show supercurrents $J_1$, $J_2$ and $J_3$ in the $\varphi_2$-$\varphi_3$ plane, which correspond to the panels (c1)-(c3). All panels have been obtained for $\gamma_1=0.45$, $\gamma_2=0.40$ and $\gamma_3=0.15$ with $U=3\Delta$, $\Gamma=\Delta$ and $\varphi_1 \equiv 0$.
		\label{fig:current}}
\end{figure}

\paragraph*{Phase diagrams:}
The introduced mapping reduces the solution of any $n$-terminal setup to that of the corresponding symmetric two-terminal configuration. Consequently, the ground state (GS) of an $n$-terminal SC-AIM is restricted to either singlet or doublet parity with a quantum phase transition (QPT) occurring at specific critical values $\chi^*$ fixed only by the combination of $U$, $\Gamma$, and $\Delta$~\cite{Meden-2019}. The phase diagram therefore turns out to be a contour-plot of $\chi$ which designates a singlet GS for $\chi>\chi^*$ and a doublet for $\chi<\chi^*$.

For instance, when we select $n=3$ with $U=3\Delta$, $\varepsilon_d=-U/2$ and $\Gamma=\Delta$, the QPT point of the corresponding symmetric two-terminal set-up, as determined by NRG, is $\chi^* \approx 0.721$ [see Fig.~\ref{fig:current}(a)]. Setting now $\varphi_1=0$, by virtue of gauge invariance, phase diagrams in the $\varphi_2-\varphi_3$ plane are obtained by contour-plotting $\chi$ as shown for varying $\gamma_2$ and $\gamma_3$ in Fig.~\ref{fig:triangle} in the left triangular map with six particular cases labeled {\bf A} to {\bf F} highlighted on its right. The critical value of $\chi^*$ is denoted in white and marks the boundary between the singlet (blue) and doublet (red) GS regions.

Notably, {\bf A} and {\bf C} correspond to the two-terminal setup with a single independent phase difference, as one of the leads is disconnected. The $\varphi_2$-$\varphi_3$ maps contain then redundant information in the form of perfect stripes of equivalued $\chi$. Adding a third weakly-coupled terminal, initially only bends the parity transition lines as showcased in the inset {\bf B}. Moving toward the center of the left triangle map, the bending intensifies, giving rise to isolated pockets of singlet GS in the middle of the phase diagrams, while remnants of the doublet stripes remain as seen in {\bf D}. A further increase of $\gamma_1$ leads then to the breakdown of the doublet GS stripes, transforming them into four doublet GS pockets at the corners of the first BZ, as depicted in {\bf E} and {\bf F}.

In the end, the universal nature of the $\chi$ factor in determining phase diagrams cannot be understated, as changing $U$ and $\Gamma$ requires just re-plotting of the above color maps by using the respective $\chi^*$. On the other hand, when $n$ is altered, one updates only the geometric factor $\chi$, but $\chi^*$ remains the same (see SM \cite{supp} for examples of $n=4$).

\paragraph*{High symmetry points:}
When $\chi=0$, a special high-symmetry point appears at $\varepsilon_d=-U/2$. Here, the doublet GS and a finite energy crossing of two excited singlets is protected by an additional symmetry in the spin space of the QD for an arbitrary number of terminals \footnote{The Hamiltonian is invariant under $d^{\vphantom{\dagger}}_{\uparrow}\rightarrow d^{\vphantom{\dagger}}_{\downarrow}$, $d^{\vphantom{\dagger}}_{\downarrow}\rightarrow d^{\vphantom{\dagger}}_{\uparrow}$ followed by a complex conjugation. Note that this antiunitary symmetry is different from time-reversal symmetry attained for two-terminal systems at $\varphi=0$ as noted in Ref.~\cite{Zitko-arxiv}.} 
as illustrated for $n=2$ in Fig.~\ref{fig:model}(c) and for $n=3$ in Fig.~\ref{fig:model}(d). Recently, it has gained a lot of attention since it relates to the so-called doublet chimney in the phase diagram of the two terminal setup~\cite{Bargerbos-2022,Zitko-arxiv, Paaske-arxiv}. However, it is crucial to note that in the  two-terminal setup, this high-symmetry point can only be realized under perfectly symmetric coupling conditions. Consequently, its experimental realization remains a formidable challenge, primarily due to the limited control over coupling strengths during device fabrication, although the doublet chimney related to the symmetry of this point is far more robust and was already realized experimentally in Ref.~\cite{Bargerbos-2022}. 

Nevertheless, in multi-terminal setups the realization of high symmetry points is more straightforward, as can be deduced from Fig.~\ref{fig:model}(b). To obtain $\chi=0$, a closed loop, which begins and ends in zero, needs to be formed when summing $\gamma_j e^{i \varphi_j}$ contributions together. Clearly, the task simplifies as $n$ increases. Already for $n=3$, the solution for the phases reads 
\begin{align}
	\varphi_j^\pm &= \pi \mp (-1)^j\arccos \frac{1-2(\gamma_2 +\gamma_3 -\gamma_2\gamma_3 +\gamma^2_j)}{2(1-\gamma_2-\gamma_3)\gamma_j} + 2\pi z_j
    \label{eq:chi_zero}
\end{align}
with $j \in 2,3$ ($\varphi_1\equiv0$) and $z_j\in \mathbb{Z}$, $\varphi_3=\varphi_2\pm\pi$ for $\gamma_2=\gamma_3=1/2$ and $\varphi_j=\pm\pi$ for $\gamma_1=\gamma_j=1/2$. Solution for $\chi=0$ thus exists, when $\gamma_{\text{max}}\leq 1/2$  where $\gamma_{\text{max}}=\text{max}(\{\gamma_j\})$. This defines a triangular region in the $\gamma_2$-$\gamma_3$ coupling space, as highlighted by the green dashed lines in Fig.~\ref{fig:triangle}. There are thus no $\chi=0$ points outside of this region (insets {\bf A} and {\bf F}), but within it  pairs of them appear in the first BZ (inset  {\bf D}). Exactly at the border, the pairs merge together, so the number of $\chi=0$  points reduces to one per the first BZ as seen by moving from {\bf D} to {\bf E} \footnote{The first BZ collapses to $1D$ in {\bf C} and the green line thus corresponds always to one and the same $\chi=0$ point.}. In clear contrast to the two-terminal setup, one fourth of the parameter space can be tuned into the desired $\chi=0$ regime, which paves the way for the experimental observation of these special points. They are additionally protected by symmetry and become isolated in $n=3$.  Additionally, the symmetry enforces crossing of two excited singlets at finite energy, and their overall behavior for $n=3$ is described by a two-dimensional Weyl Hamiltonian [Fig.~\ref{fig:model}(d)]. As explained in SM \cite{supp}, for $n=4$, lines or loops of $\chi=0$ points form in the three-dimensional synthetic BZ with isolated points appearing only for $\text{max}(\{\gamma_j\}) =1/2$.

\paragraph*{Josephson currents:} The versatility of multi-terminal Josephson junctions arises from their ability to divide supercurrents into $n \geq 3$ terminals, each denoted as $J_j$ and passing through the respective lead $j$, as depicted in Fig.~\ref{fig:current}(b). Diverse novel phenomena unfold then, encompassing SC transistor and SC diode effects, as well as current modulation. To demonstrate these, we apply the Hellmann-Feynman theorem at zero temperature in conjunction with the herein discovered mapping (see also SM \cite{supp}):
\begin{align}
J_j
=
\frac{2e}{\hbar} 
\frac{\partial E}{\partial\varphi_j} 
= 
\frac{2e}{\hbar} 
\frac{ \partial E}{\partial\chi}
\frac{\partial \chi}{\partial \varphi_j} 
= \mathcal{J}(\chi)
\frac{\partial \chi}{\partial \varphi_j},
\label{eq:current}
\end{align}
where $E$ is the GS energy, $\mathcal{J}(\chi)\equiv \frac{2e}{\hbar} \frac{\partial E}{\partial\chi}$ is a universal function of $\chi$ with parametric dependency on $U$ and $\Gamma$, while all geometric details, i.e.~$n$, $\gamma_j$ and $\varphi_j$, are confined only to the analytic factors $\frac{\partial \chi}{\partial\varphi_j}$. Again, the $\mathcal{J}(\chi)$ input can be extracted from the symmetric two-terminal configuration by a numeric method of choice (for details see SM \cite{supp}), with an NRG solution shown in Fig.~\ref{fig:current}(a) for $U=3\Delta$ and $\Gamma=\Delta$.

Using $\mathcal{J}(\chi)$ from Fig.~\ref{fig:current}(a), we select $\gamma_1=0.45$, $\gamma_2=0.40$ and $\gamma_3=0.15$, and determine the corresponding currents $J_j$ in the $\varphi_2$-$\varphi_3$ plane ($\varphi_1 \equiv 0$) as plotted in Figs.~\ref{fig:current}(d1)-(d3). Notably, QPTs appear congruently at elliptic-like lines dividing positive and negative values of $J_j$. When changing $n$, only the $\frac{\partial \chi}{\partial\varphi_j}$ function is updated, so the current is re-plotted correspondingly. $\mathcal{J}(\chi)$ has to be recalculated only if $U$ and/or $\Gamma$ are changed.

The parameters of resulting device allow treating the third terminal as its base, while $J_1$ and $-J_2$ are assigned to represent the input and output current, respectively. The selected signs reflect their directions according to Fig.~\ref{fig:current}(b). Keeping first $\varphi_3=0$, the resulting current phase relation (CPR) of $-J_2$, shown in Fig.~\ref{fig:current}(c1), exhibits nodes at $\varphi_2=0,\pi$ in the first BZ and additional QPT points coinciding with the elliptic-like transition lines of Fig.~\ref{fig:current}(d2). Increasing $\varphi_3$ up to $\pi/2$, the CPR is only modulated in phase (the nodes shift) and amplitude. Finally, setting $\varphi_3=\pi$ forces the device to reside exclusively in the singlet GS, which significantly suppresses $-J_2$, as seen in Fig.~\ref{fig:current}(c2). This effectively switches off the three-terminal device, akin to a SC-transistor effect. 

More elaborate phase control is possible by sweeping $\varphi_2$ and simultaneously adjusting $\varphi_3=\varphi_2+\phi$. For $\phi=\pi/2$, as depicted in Fig.~\ref{fig:current}(c3), the CPR is then tuned into a directional regime with different positive and negative critical currents. This results in a SC-diode effect with a yield of $\approx 50\%$. We emphasize that CPR modulation and diode effect are quite ubiquitous, while demonstrating the transistor effect required a weakly coupled third terminal and closed pockets of doublet GS pockets that form only at $\gamma_3 \approx 0.15$.

\paragraph*{Conclusions:}
In this letter, we explore $n$-terminal JJs based on a single-level interacting quantum dot described via SC-AIM. Our key insight is their analytic mapping onto two-terminal junctions with symmetric couplings and phase difference expressed via a single analytic quantity $\chi$. This facilitates the derivation of complete phase diagrams and associated Josephson currents, requiring only universal values of critical $\chi^*$ and the function $\mathcal{J}(\chi)$, respectively. Geometric details, including the number of leads, SC phases $\varphi_j$, and the distribution of total tunneling strength $\Gamma$ among the leads, are then fully encoded through analytic functions. Importantly, our system supports high-symmetry points, where doublet GS and finite energy band crossing are protected, within a substantial region of experimentally well-accessible coupling space unlike its two-terminal counterpart. In addition, our research emphasizes the practical advantages of integrating three-terminal quantum dot-based devices into Josephson junction circuits to leverage their transistor or diode effects. 

\begin{acknowledgements}
	The authors would like to thank Virgil V. Baran, Wolfgang Belzig, Jens Paaske and Rok \v{Z}itko for helpful discussions. This work was supported by Grant No. 23-05263K of the Czech Science Foundation and the project e-INFRA CZ (ID:90140) of the Czech Ministry of Education, Youth and Sports.
\end{acknowledgements}


\begin{thebibliography}{53}%
	\makeatletter
	\providecommand \@ifxundefined [1]{%
		\@ifx{#1\undefined}
	}%
	\providecommand \@ifnum [1]{%
		\ifnum #1\expandafter \@firstoftwo
		\else \expandafter \@secondoftwo
		\fi
	}%
	\providecommand \@ifx [1]{%
		\ifx #1\expandafter \@firstoftwo
		\else \expandafter \@secondoftwo
		\fi
	}%
	\providecommand \natexlab [1]{#1}%
	\providecommand \enquote  [1]{``#1''}%
	\providecommand \bibnamefont  [1]{#1}%
	\providecommand \bibfnamefont [1]{#1}%
	\providecommand \citenamefont [1]{#1}%
	\providecommand \href@noop [0]{\@secondoftwo}%
	\providecommand \href [0]{\begingroup \@sanitize@url \@href}%
	\providecommand \@href[1]{\@@startlink{#1}\@@href}%
	\providecommand \@@href[1]{\endgroup#1\@@endlink}%
	\providecommand \@sanitize@url [0]{\catcode `\\12\catcode `\$12\catcode
		`\&12\catcode `\#12\catcode `\^12\catcode `\_12\catcode `\%12\relax}%
	\providecommand \@@startlink[1]{}%
	\providecommand \@@endlink[0]{}%
	\providecommand \url  [0]{\begingroup\@sanitize@url \@url }%
	\providecommand \@url [1]{\endgroup\@href {#1}{\urlprefix }}%
	\providecommand \urlprefix  [0]{URL }%
	\providecommand \Eprint [0]{\href }%
	\providecommand \doibase [0]{https://doi.org/}%
	\providecommand \selectlanguage [0]{\@gobble}%
	\providecommand \bibinfo  [0]{\@secondoftwo}%
	\providecommand \bibfield  [0]{\@secondoftwo}%
	\providecommand \translation [1]{[#1]}%
	\providecommand \BibitemOpen [0]{}%
	\providecommand \bibitemStop [0]{}%
	\providecommand \bibitemNoStop [0]{.\EOS\space}%
	\providecommand \EOS [0]{\spacefactor3000\relax}%
	\providecommand \BibitemShut  [1]{\csname bibitem#1\endcsname}%
	\let\auto@bib@innerbib\@empty
	\bibitem [{\citenamefont {Likharev}\ and\ \citenamefont
		{Semenov}(1991)}]{Likharev-1991}%
	\BibitemOpen
	\bibfield  {author} {\bibinfo {author} {\bibfnamefont {K.}~\bibnamefont
			{Likharev}}\ and\ \bibinfo {author} {\bibfnamefont {V.}~\bibnamefont
			{Semenov}},\ }\href {https://doi.org/10.1109/77.80745} {\bibfield  {journal}
		{\bibinfo  {journal} {IEEE Transactions on Applied Superconductivity}\
		}\textbf {\bibinfo {volume} {1}},\ \bibinfo {pages} {3} (\bibinfo {year}
		{1991})}\BibitemShut {NoStop}%
	\bibitem [{\citenamefont {Cleuziou}\ \emph {et~al.}(2006)\citenamefont
		{Cleuziou}, \citenamefont {Wernsdorfer}, \citenamefont {Bouchiat},
		\citenamefont {Ondarcuhu},\ and\ \citenamefont {Monthioux}}]{Cleuziou-2006}%
	\BibitemOpen
	\bibfield  {author} {\bibinfo {author} {\bibfnamefont {J.~P.}\ \bibnamefont
			{Cleuziou}}, \bibinfo {author} {\bibfnamefont {W.}~\bibnamefont
			{Wernsdorfer}}, \bibinfo {author} {\bibfnamefont {V.}~\bibnamefont
			{Bouchiat}}, \bibinfo {author} {\bibfnamefont {T.}~\bibnamefont
			{Ondarcuhu}},\ and\ \bibinfo {author} {\bibfnamefont {M.}~\bibnamefont
			{Monthioux}},\ }\href {http://dx.doi.org/10.1038/nnano.2006.54} {\bibfield
		{journal} {\bibinfo  {journal} {Nat. Nanotechnol.}\ }\textbf {\bibinfo
			{volume} {1}},\ \bibinfo {pages} {53} (\bibinfo {year} {2006})}\BibitemShut
	{NoStop}%
	\bibitem [{\citenamefont {Clarke}\ and\ \citenamefont
		{Wilhelm}(2008)}]{Clarke-2008}%
	\BibitemOpen
	\bibfield  {author} {\bibinfo {author} {\bibfnamefont {J.}~\bibnamefont
			{Clarke}}\ and\ \bibinfo {author} {\bibfnamefont {F.~K.}\ \bibnamefont
			{Wilhelm}},\ }\href {https://doi.org/10.1038/nature07128} {\bibfield
		{journal} {\bibinfo  {journal} {Nature}\ }\textbf {\bibinfo {volume} {453}},\
		\bibinfo {pages} {1031} (\bibinfo {year} {2008})}\BibitemShut {NoStop}%
	\bibitem [{\citenamefont {Mart{\'\i}n-Rodero}\ and\ \citenamefont
		{Levy~Yeyati}(2011)}]{Rodero-2011}%
	\BibitemOpen
	\bibfield  {author} {\bibinfo {author} {\bibfnamefont {A.}~\bibnamefont
			{Mart{\'\i}n-Rodero}}\ and\ \bibinfo {author} {\bibfnamefont
			{A.}~\bibnamefont {Levy~Yeyati}},\ }\href
	{https://doi.org/10.1080/00018732.2011.624266} {\bibfield  {journal}
		{\bibinfo  {journal} {Adv. Phys.}\ }\textbf {\bibinfo {volume} {60}},\
		\bibinfo {pages} {899} (\bibinfo {year} {2011})}\BibitemShut {NoStop}%
	\bibitem [{\citenamefont {Meden}(2019)}]{Meden-2019}%
	\BibitemOpen
	\bibfield  {author} {\bibinfo {author} {\bibfnamefont {V.}~\bibnamefont
			{Meden}},\ }\href {https://doi.org/10.1088/1361-648x/aafd6a} {\bibfield
		{journal} {\bibinfo  {journal} {Journal of Physics: Condensed Matter}\
		}\textbf {\bibinfo {volume} {31}},\ \bibinfo {pages} {163001} (\bibinfo
		{year} {2019})}\BibitemShut {NoStop}%
	\bibitem [{\citenamefont {van Heck}\ \emph {et~al.}(2014)\citenamefont {van
			Heck}, \citenamefont {Mi},\ and\ \citenamefont {Akhmerov}}]{Heck-2014}%
	\BibitemOpen
	\bibfield  {author} {\bibinfo {author} {\bibfnamefont {B.}~\bibnamefont {van
				Heck}}, \bibinfo {author} {\bibfnamefont {S.}~\bibnamefont {Mi}},\ and\
		\bibinfo {author} {\bibfnamefont {A.~R.}\ \bibnamefont {Akhmerov}},\ }\href
	{https://doi.org/10.1103/PhysRevB.90.155450} {\bibfield  {journal} {\bibinfo
			{journal} {Phys. Rev. B}\ }\textbf {\bibinfo {volume} {90}},\ \bibinfo
		{pages} {155450} (\bibinfo {year} {2014})}\BibitemShut {NoStop}%
	\bibitem [{\citenamefont {Riwar}\ \emph {et~al.}(2016)\citenamefont {Riwar},
		\citenamefont {Houzet}, \citenamefont {Meyer},\ and\ \citenamefont
		{Nazarov}}]{Riwar-2016}%
	\BibitemOpen
	\bibfield  {author} {\bibinfo {author} {\bibfnamefont {R.-P.}\ \bibnamefont
			{Riwar}}, \bibinfo {author} {\bibfnamefont {M.}~\bibnamefont {Houzet}},
		\bibinfo {author} {\bibfnamefont {J.~S.}\ \bibnamefont {Meyer}},\ and\
		\bibinfo {author} {\bibfnamefont {Y.~V.}\ \bibnamefont {Nazarov}},\ }\href
	{https://doi.org/10.1038/ncomms11167} {\bibfield  {journal} {\bibinfo
			{journal} {Nature Communications}\ }\textbf {\bibinfo {volume} {7}},\
		\bibinfo {pages} {11167} (\bibinfo {year} {2016})}\BibitemShut {NoStop}%
	\bibitem [{\citenamefont {Eriksson}\ \emph {et~al.}(2017)\citenamefont
		{Eriksson}, \citenamefont {Riwar}, \citenamefont {Houzet}, \citenamefont
		{Meyer},\ and\ \citenamefont {Nazarov}}]{Eriksson-2017}%
	\BibitemOpen
	\bibfield  {author} {\bibinfo {author} {\bibfnamefont {E.}~\bibnamefont
			{Eriksson}}, \bibinfo {author} {\bibfnamefont {R.-P.}\ \bibnamefont {Riwar}},
		\bibinfo {author} {\bibfnamefont {M.}~\bibnamefont {Houzet}}, \bibinfo
		{author} {\bibfnamefont {J.~S.}\ \bibnamefont {Meyer}},\ and\ \bibinfo
		{author} {\bibfnamefont {Y.~V.}\ \bibnamefont {Nazarov}},\ }\href
	{https://doi.org/10.1103/PhysRevB.95.075417} {\bibfield  {journal} {\bibinfo
			{journal} {Phys. Rev. B}\ }\textbf {\bibinfo {volume} {95}},\ \bibinfo
		{pages} {075417} (\bibinfo {year} {2017})}\BibitemShut {NoStop}%
	\bibitem [{\citenamefont {Lutchyn}\ \emph {et~al.}(2018)\citenamefont
		{Lutchyn}, \citenamefont {Bakkers}, \citenamefont {Kouwenhoven},
		\citenamefont {Krogstrup}, \citenamefont {Marcus},\ and\ \citenamefont
		{Oreg}}]{Lutchyn-2018}%
	\BibitemOpen
	\bibfield  {author} {\bibinfo {author} {\bibfnamefont {R.~M.}\ \bibnamefont
			{Lutchyn}}, \bibinfo {author} {\bibfnamefont {E.~P. A.~M.}\ \bibnamefont
			{Bakkers}}, \bibinfo {author} {\bibfnamefont {L.~P.}\ \bibnamefont
			{Kouwenhoven}}, \bibinfo {author} {\bibfnamefont {P.}~\bibnamefont
			{Krogstrup}}, \bibinfo {author} {\bibfnamefont {C.~M.}\ \bibnamefont
			{Marcus}},\ and\ \bibinfo {author} {\bibfnamefont {Y.}~\bibnamefont {Oreg}},\
	}\href {https://doi.org/10.1038/s41578-018-0003-1} {\bibfield  {journal}
		{\bibinfo  {journal} {Nature Reviews Materials}\ }\textbf {\bibinfo {volume}
			{3}},\ \bibinfo {pages} {52} (\bibinfo {year} {2018})}\BibitemShut {NoStop}%
	\bibitem [{\citenamefont {Xie}\ \emph {et~al.}(2018)\citenamefont {Xie},
		\citenamefont {Vavilov},\ and\ \citenamefont {Levchenko}}]{Xie-2018}%
	\BibitemOpen
	\bibfield  {author} {\bibinfo {author} {\bibfnamefont {H.-Y.}\ \bibnamefont
			{Xie}}, \bibinfo {author} {\bibfnamefont {M.~G.}\ \bibnamefont {Vavilov}},\
		and\ \bibinfo {author} {\bibfnamefont {A.}~\bibnamefont {Levchenko}},\ }\href
	{https://doi.org/10.1103/PhysRevB.97.035443} {\bibfield  {journal} {\bibinfo
			{journal} {Phys. Rev. B}\ }\textbf {\bibinfo {volume} {97}},\ \bibinfo
		{pages} {035443} (\bibinfo {year} {2018})}\BibitemShut {NoStop}%
	\bibitem [{\citenamefont {Xie}\ and\ \citenamefont
		{Levchenko}(2019)}]{Xie-2019}%
	\BibitemOpen
	\bibfield  {author} {\bibinfo {author} {\bibfnamefont {H.-Y.}\ \bibnamefont
			{Xie}}\ and\ \bibinfo {author} {\bibfnamefont {A.}~\bibnamefont
			{Levchenko}},\ }\href {https://doi.org/10.1103/PhysRevB.99.094519} {\bibfield
		{journal} {\bibinfo  {journal} {Phys. Rev. B}\ }\textbf {\bibinfo {volume}
			{99}},\ \bibinfo {pages} {094519} (\bibinfo {year} {2019})}\BibitemShut
	{NoStop}%
	\bibitem [{\citenamefont {Repin}\ and\ \citenamefont
		{Nazarov}(2022)}]{Repin-2022}%
	\BibitemOpen
	\bibfield  {author} {\bibinfo {author} {\bibfnamefont {E.~V.}\ \bibnamefont
			{Repin}}\ and\ \bibinfo {author} {\bibfnamefont {Y.~V.}\ \bibnamefont
			{Nazarov}},\ }\href {https://doi.org/10.1103/PhysRevB.105.L041405} {\bibfield
		{journal} {\bibinfo  {journal} {Phys. Rev. B}\ }\textbf {\bibinfo {volume}
			{105}},\ \bibinfo {pages} {L041405} (\bibinfo {year} {2022})}\BibitemShut
	{NoStop}%
	\bibitem [{\citenamefont {Gavensky}\ \emph {et~al.}(2023)\citenamefont
		{Gavensky}, \citenamefont {Usaj},\ and\ \citenamefont
		{Balseiro}}]{Gavensky-2023}%
	\BibitemOpen
	\bibfield  {author} {\bibinfo {author} {\bibfnamefont {L.~P.}\ \bibnamefont
			{Gavensky}}, \bibinfo {author} {\bibfnamefont {G.}~\bibnamefont {Usaj}},\
		and\ \bibinfo {author} {\bibfnamefont {C.~A.}\ \bibnamefont {Balseiro}},\
	}\href {https://doi.org/10.1209/0295-5075/acb2f6} {\bibfield  {journal}
		{\bibinfo  {journal} {Europhysics Letters}\ }\textbf {\bibinfo {volume}
			{141}},\ \bibinfo {pages} {36001} (\bibinfo {year} {2023})}\BibitemShut
	{NoStop}%
	\bibitem [{\citenamefont {Alicea}\ \emph {et~al.}(2011)\citenamefont {Alicea},
		\citenamefont {Oreg}, \citenamefont {Refael}, \citenamefont {von Oppen},\
		and\ \citenamefont {Fisher}}]{Alicea-2011}%
	\BibitemOpen
	\bibfield  {author} {\bibinfo {author} {\bibfnamefont {J.}~\bibnamefont
			{Alicea}}, \bibinfo {author} {\bibfnamefont {Y.}~\bibnamefont {Oreg}},
		\bibinfo {author} {\bibfnamefont {G.}~\bibnamefont {Refael}}, \bibinfo
		{author} {\bibfnamefont {F.}~\bibnamefont {von Oppen}},\ and\ \bibinfo
		{author} {\bibfnamefont {M.~P.~A.}\ \bibnamefont {Fisher}},\ }\href
	{https://doi.org/10.1038/nphys1915} {\bibfield  {journal} {\bibinfo
			{journal} {Nature Physics}\ }\textbf {\bibinfo {volume} {7}},\ \bibinfo
		{pages} {412} (\bibinfo {year} {2011})}\BibitemShut {NoStop}%
	\bibitem [{\citenamefont {Zazunov}\ \emph {et~al.}(2017)\citenamefont
		{Zazunov}, \citenamefont {Egger}, \citenamefont {Alvarado},\ and\
		\citenamefont {Yeyati}}]{Zazunov-2017}%
	\BibitemOpen
	\bibfield  {author} {\bibinfo {author} {\bibfnamefont {A.}~\bibnamefont
			{Zazunov}}, \bibinfo {author} {\bibfnamefont {R.}~\bibnamefont {Egger}},
		\bibinfo {author} {\bibfnamefont {M.}~\bibnamefont {Alvarado}},\ and\
		\bibinfo {author} {\bibfnamefont {A.~L.}\ \bibnamefont {Yeyati}},\ }\href
	{https://doi.org/10.1103/PhysRevB.96.024516} {\bibfield  {journal} {\bibinfo
			{journal} {Phys. Rev. B}\ }\textbf {\bibinfo {volume} {96}},\ \bibinfo
		{pages} {024516} (\bibinfo {year} {2017})}\BibitemShut {NoStop}%
	\bibitem [{\citenamefont {Freyn}\ \emph {et~al.}(2011)\citenamefont {Freyn},
		\citenamefont {Dou\ifmmode~\mbox{\c{c}}\else \c{c}\fi{}ot}, \citenamefont
		{Feinberg},\ and\ \citenamefont {M\'elin}}]{Freyn-2011}%
	\BibitemOpen
	\bibfield  {author} {\bibinfo {author} {\bibfnamefont {A.}~\bibnamefont
			{Freyn}}, \bibinfo {author} {\bibfnamefont {B.}~\bibnamefont
			{Dou\ifmmode~\mbox{\c{c}}\else \c{c}\fi{}ot}}, \bibinfo {author}
		{\bibfnamefont {D.}~\bibnamefont {Feinberg}},\ and\ \bibinfo {author}
		{\bibfnamefont {R.}~\bibnamefont {M\'elin}},\ }\href
	{https://doi.org/10.1103/PhysRevLett.106.257005} {\bibfield  {journal}
		{\bibinfo  {journal} {Phys. Rev. Lett.}\ }\textbf {\bibinfo {volume} {106}},\
		\bibinfo {pages} {257005} (\bibinfo {year} {2011})}\BibitemShut {NoStop}%
	\bibitem [{\citenamefont {Draelos}\ \emph {et~al.}(2019)\citenamefont
		{Draelos}, \citenamefont {Wei}, \citenamefont {Seredinski}, \citenamefont
		{Li}, \citenamefont {Mehta}, \citenamefont {Watanabe}, \citenamefont
		{Taniguchi}, \citenamefont {Borzenets}, \citenamefont {Amet},\ and\
		\citenamefont {Finkelstein}}]{Draelos-2019}%
	\BibitemOpen
	\bibfield  {author} {\bibinfo {author} {\bibfnamefont {A.~W.}\ \bibnamefont
			{Draelos}}, \bibinfo {author} {\bibfnamefont {M.-T.}\ \bibnamefont {Wei}},
		\bibinfo {author} {\bibfnamefont {A.}~\bibnamefont {Seredinski}}, \bibinfo
		{author} {\bibfnamefont {H.}~\bibnamefont {Li}}, \bibinfo {author}
		{\bibfnamefont {Y.}~\bibnamefont {Mehta}}, \bibinfo {author} {\bibfnamefont
			{K.}~\bibnamefont {Watanabe}}, \bibinfo {author} {\bibfnamefont
			{T.}~\bibnamefont {Taniguchi}}, \bibinfo {author} {\bibfnamefont {I.~V.}\
			\bibnamefont {Borzenets}}, \bibinfo {author} {\bibfnamefont {F.}~\bibnamefont
			{Amet}},\ and\ \bibinfo {author} {\bibfnamefont {G.}~\bibnamefont
			{Finkelstein}},\ }\href {https://doi.org/10.1021/acs.nanolett.8b04330}
	{\bibfield  {journal} {\bibinfo  {journal} {Nano Letters}\ }\textbf {\bibinfo
			{volume} {19}},\ \bibinfo {pages} {1039} (\bibinfo {year}
		{2019})}\BibitemShut {NoStop}%
	\bibitem [{\citenamefont {Pankratova}\ \emph {et~al.}(2020)\citenamefont
		{Pankratova}, \citenamefont {Lee}, \citenamefont {Kuzmin}, \citenamefont
		{Wickramasinghe}, \citenamefont {Mayer}, \citenamefont {Yuan}, \citenamefont
		{Vavilov}, \citenamefont {Shabani},\ and\ \citenamefont
		{Manucharyan}}]{Pankratova-2020}%
	\BibitemOpen
	\bibfield  {author} {\bibinfo {author} {\bibfnamefont {N.}~\bibnamefont
			{Pankratova}}, \bibinfo {author} {\bibfnamefont {H.}~\bibnamefont {Lee}},
		\bibinfo {author} {\bibfnamefont {R.}~\bibnamefont {Kuzmin}}, \bibinfo
		{author} {\bibfnamefont {K.}~\bibnamefont {Wickramasinghe}}, \bibinfo
		{author} {\bibfnamefont {W.}~\bibnamefont {Mayer}}, \bibinfo {author}
		{\bibfnamefont {J.}~\bibnamefont {Yuan}}, \bibinfo {author} {\bibfnamefont
			{M.~G.}\ \bibnamefont {Vavilov}}, \bibinfo {author} {\bibfnamefont
			{J.}~\bibnamefont {Shabani}},\ and\ \bibinfo {author} {\bibfnamefont {V.~E.}\
			\bibnamefont {Manucharyan}},\ }\href
	{https://doi.org/10.1103/PhysRevX.10.031051} {\bibfield  {journal} {\bibinfo
			{journal} {Phys. Rev. X}\ }\textbf {\bibinfo {volume} {10}},\ \bibinfo
		{pages} {031051} (\bibinfo {year} {2020})}\BibitemShut {NoStop}%
	\bibitem [{\citenamefont {Graziano}\ \emph {et~al.}(2022)\citenamefont
		{Graziano}, \citenamefont {Gupta}, \citenamefont {Pendharkar}, \citenamefont
		{Dong}, \citenamefont {Dempsey}, \citenamefont {Palmstr{\o}m},\ and\
		\citenamefont {Pribiag}}]{Graziano-2022}%
	\BibitemOpen
	\bibfield  {author} {\bibinfo {author} {\bibfnamefont {G.~V.}\ \bibnamefont
			{Graziano}}, \bibinfo {author} {\bibfnamefont {M.}~\bibnamefont {Gupta}},
		\bibinfo {author} {\bibfnamefont {M.}~\bibnamefont {Pendharkar}}, \bibinfo
		{author} {\bibfnamefont {J.~T.}\ \bibnamefont {Dong}}, \bibinfo {author}
		{\bibfnamefont {C.~P.}\ \bibnamefont {Dempsey}}, \bibinfo {author}
		{\bibfnamefont {C.}~\bibnamefont {Palmstr{\o}m}},\ and\ \bibinfo {author}
		{\bibfnamefont {V.~S.}\ \bibnamefont {Pribiag}},\ }\href
	{https://doi.org/10.1038/s41467-022-33682-2} {\bibfield  {journal} {\bibinfo
			{journal} {Nature Communications}\ }\textbf {\bibinfo {volume} {13}},\
		\bibinfo {pages} {5933} (\bibinfo {year} {2022})}\BibitemShut {NoStop}%
	\bibitem [{\citenamefont {Gupta}\ \emph {et~al.}(2023)\citenamefont {Gupta},
		\citenamefont {Graziano}, \citenamefont {Pendharkar}, \citenamefont {Dong},
		\citenamefont {Dempsey}, \citenamefont {Palmstr{\o}m},\ and\ \citenamefont
		{Pribiag}}]{Gupta-2023}%
	\BibitemOpen
	\bibfield  {author} {\bibinfo {author} {\bibfnamefont {M.}~\bibnamefont
			{Gupta}}, \bibinfo {author} {\bibfnamefont {G.~V.}\ \bibnamefont {Graziano}},
		\bibinfo {author} {\bibfnamefont {M.}~\bibnamefont {Pendharkar}}, \bibinfo
		{author} {\bibfnamefont {J.~T.}\ \bibnamefont {Dong}}, \bibinfo {author}
		{\bibfnamefont {C.~P.}\ \bibnamefont {Dempsey}}, \bibinfo {author}
		{\bibfnamefont {C.}~\bibnamefont {Palmstr{\o}m}},\ and\ \bibinfo {author}
		{\bibfnamefont {V.~S.}\ \bibnamefont {Pribiag}},\ }\href
	{https://doi.org/10.1038/s41467-023-38856-0} {\bibfield  {journal} {\bibinfo
			{journal} {Nature Communications}\ }\textbf {\bibinfo {volume} {14}},\
		\bibinfo {pages} {3078} (\bibinfo {year} {2023})}\BibitemShut {NoStop}%
	\bibitem [{\citenamefont {M\'elin}\ and\ \citenamefont
		{Feinberg}(2023)}]{Melin-2023}%
	\BibitemOpen
	\bibfield  {author} {\bibinfo {author} {\bibfnamefont {R.}~\bibnamefont
			{M\'elin}}\ and\ \bibinfo {author} {\bibfnamefont {D.}~\bibnamefont
			{Feinberg}},\ }\href {https://doi.org/10.1103/PhysRevB.107.L161405}
	{\bibfield  {journal} {\bibinfo  {journal} {Phys. Rev. B}\ }\textbf {\bibinfo
			{volume} {107}},\ \bibinfo {pages} {L161405} (\bibinfo {year}
		{2023})}\BibitemShut {NoStop}%
	\bibitem [{\citenamefont {Zhang}\ \emph {et~al.}(2023)\citenamefont {Zhang},
		\citenamefont {Rashid}, \citenamefont {Ahari}, \citenamefont {Zhang},
		\citenamefont {Ananthanarayanan}, \citenamefont {Xiao}, \citenamefont
		{de~Coster}, \citenamefont {Gilbert}, \citenamefont {Samarth},\ and\
		\citenamefont {Kayyalha}}]{Zhang-2023}%
	\BibitemOpen
	\bibfield  {author} {\bibinfo {author} {\bibfnamefont {F.}~\bibnamefont
			{Zhang}}, \bibinfo {author} {\bibfnamefont {A.~S.}\ \bibnamefont {Rashid}},
		\bibinfo {author} {\bibfnamefont {M.~T.}\ \bibnamefont {Ahari}}, \bibinfo
		{author} {\bibfnamefont {W.}~\bibnamefont {Zhang}}, \bibinfo {author}
		{\bibfnamefont {K.~M.}\ \bibnamefont {Ananthanarayanan}}, \bibinfo {author}
		{\bibfnamefont {R.}~\bibnamefont {Xiao}}, \bibinfo {author} {\bibfnamefont
			{G.~J.}\ \bibnamefont {de~Coster}}, \bibinfo {author} {\bibfnamefont {M.~J.}\
			\bibnamefont {Gilbert}}, \bibinfo {author} {\bibfnamefont {N.}~\bibnamefont
			{Samarth}},\ and\ \bibinfo {author} {\bibfnamefont {M.}~\bibnamefont
			{Kayyalha}},\ }\href {https://doi.org/10.1103/PhysRevB.107.L140503}
	{\bibfield  {journal} {\bibinfo  {journal} {Phys. Rev. B}\ }\textbf {\bibinfo
			{volume} {107}},\ \bibinfo {pages} {L140503} (\bibinfo {year}
		{2023})}\BibitemShut {NoStop}%
	\bibitem [{\citenamefont {Coraiola}\ \emph {et~al.}(2023)\citenamefont
		{Coraiola}, \citenamefont {Haxell}, \citenamefont {Sabonis}, \citenamefont
		{Weisbrich}, \citenamefont {Svetogorov}, \citenamefont {Hinderling},
		\citenamefont {ten Kate}, \citenamefont {Cheah}, \citenamefont {Krizek},
		\citenamefont {Schott}, \citenamefont {Wegscheider}, \citenamefont {Cuevas},
		\citenamefont {Belzig},\ and\ \citenamefont {Nichele}}]{Coraiola-2023}%
	\BibitemOpen
	\bibfield  {author} {\bibinfo {author} {\bibfnamefont {M.}~\bibnamefont
			{Coraiola}}, \bibinfo {author} {\bibfnamefont {D.~Z.}\ \bibnamefont
			{Haxell}}, \bibinfo {author} {\bibfnamefont {D.}~\bibnamefont {Sabonis}},
		\bibinfo {author} {\bibfnamefont {H.}~\bibnamefont {Weisbrich}}, \bibinfo
		{author} {\bibfnamefont {A.~E.}\ \bibnamefont {Svetogorov}}, \bibinfo
		{author} {\bibfnamefont {M.}~\bibnamefont {Hinderling}}, \bibinfo {author}
		{\bibfnamefont {S.~C.}\ \bibnamefont {ten Kate}}, \bibinfo {author}
		{\bibfnamefont {E.}~\bibnamefont {Cheah}}, \bibinfo {author} {\bibfnamefont
			{F.}~\bibnamefont {Krizek}}, \bibinfo {author} {\bibfnamefont
			{R.}~\bibnamefont {Schott}}, \bibinfo {author} {\bibfnamefont
			{W.}~\bibnamefont {Wegscheider}}, \bibinfo {author} {\bibfnamefont {J.~C.}\
			\bibnamefont {Cuevas}}, \bibinfo {author} {\bibfnamefont {W.}~\bibnamefont
			{Belzig}},\ and\ \bibinfo {author} {\bibfnamefont {F.}~\bibnamefont
			{Nichele}},\ }\href@noop {} {\bibinfo {title} {Hybridisation of andreev bound
			states in three-terminal josephson junctions}} (\bibinfo {year} {2023}),\
	\Eprint {https://arxiv.org/abs/2302.14535} {arXiv:2302.14535
		[cond-mat.mes-hall]} \BibitemShut {NoStop}%
	\bibitem [{\citenamefont {Matsuo}\ \emph {et~al.}(2023)\citenamefont {Matsuo},
		\citenamefont {Imoto}, \citenamefont {Yokoyama}, \citenamefont {Sato},
		\citenamefont {Lindemann}, \citenamefont {Gronin}, \citenamefont {Gardner},
		\citenamefont {Nakosai}, \citenamefont {Tanaka}, \citenamefont {Manfra},\
		and\ \citenamefont {Tarucha}}]{Sadashige-2023}%
	\BibitemOpen
	\bibfield  {author} {\bibinfo {author} {\bibfnamefont {S.}~\bibnamefont
			{Matsuo}}, \bibinfo {author} {\bibfnamefont {T.}~\bibnamefont {Imoto}},
		\bibinfo {author} {\bibfnamefont {T.}~\bibnamefont {Yokoyama}}, \bibinfo
		{author} {\bibfnamefont {Y.}~\bibnamefont {Sato}}, \bibinfo {author}
		{\bibfnamefont {T.}~\bibnamefont {Lindemann}}, \bibinfo {author}
		{\bibfnamefont {S.}~\bibnamefont {Gronin}}, \bibinfo {author} {\bibfnamefont
			{G.~C.}\ \bibnamefont {Gardner}}, \bibinfo {author} {\bibfnamefont
			{S.}~\bibnamefont {Nakosai}}, \bibinfo {author} {\bibfnamefont
			{Y.}~\bibnamefont {Tanaka}}, \bibinfo {author} {\bibfnamefont {M.~J.}\
			\bibnamefont {Manfra}},\ and\ \bibinfo {author} {\bibfnamefont
			{S.}~\bibnamefont {Tarucha}},\ }\href@noop {} {\bibinfo {title}
		{Phase-dependent andreev molecules and superconducting gap closing in
			coherently coupled josephson junctions}} (\bibinfo {year} {2023}),\ \Eprint
	{https://arxiv.org/abs/2303.10540} {arXiv:2303.10540 [cond-mat.supr-con]}
	\BibitemShut {NoStop}%
	\bibitem [{\citenamefont {Meng}\ \emph {et~al.}(2009)\citenamefont {Meng},
		\citenamefont {Florens},\ and\ \citenamefont {Simon}}]{Meng-2009}%
	\BibitemOpen
	\bibfield  {author} {\bibinfo {author} {\bibfnamefont {T.}~\bibnamefont
			{Meng}}, \bibinfo {author} {\bibfnamefont {S.}~\bibnamefont {Florens}},\ and\
		\bibinfo {author} {\bibfnamefont {P.}~\bibnamefont {Simon}},\ }\href
	{http://link.aps.org/doi/10.1103/PhysRevB.79.224521} {\bibfield  {journal}
		{\bibinfo  {journal} {Phys. Rev. B}\ }\textbf {\bibinfo {volume} {79}},\
		\bibinfo {pages} {224521} (\bibinfo {year} {2009})}\BibitemShut {NoStop}%
	\bibitem [{\citenamefont {Grove-Rasmussen}\ \emph {et~al.}(2018)\citenamefont
		{Grove-Rasmussen}, \citenamefont {Steffensen}, \citenamefont {Jellinggaard},
		\citenamefont {Madsen}, \citenamefont {{\v{Z}}itko}, \citenamefont {Paaske},\
		and\ \citenamefont {Nyg{\aa}rd}}]{Grove-Rasmussen-2018}%
	\BibitemOpen
	\bibfield  {author} {\bibinfo {author} {\bibfnamefont {K.}~\bibnamefont
			{Grove-Rasmussen}}, \bibinfo {author} {\bibfnamefont {G.}~\bibnamefont
			{Steffensen}}, \bibinfo {author} {\bibfnamefont {A.}~\bibnamefont
			{Jellinggaard}}, \bibinfo {author} {\bibfnamefont {M.~H.}\ \bibnamefont
			{Madsen}}, \bibinfo {author} {\bibfnamefont {R.}~\bibnamefont {{\v{Z}}itko}},
		\bibinfo {author} {\bibfnamefont {J.}~\bibnamefont {Paaske}},\ and\ \bibinfo
		{author} {\bibfnamefont {J.}~\bibnamefont {Nyg{\aa}rd}},\ }\href
	{https://doi.org/10.1038/s41467-018-04683-x} {\bibfield  {journal} {\bibinfo
			{journal} {Nature Communications}\ }\textbf {\bibinfo {volume} {9}},\
		\bibinfo {pages} {2376} (\bibinfo {year} {2018})}\BibitemShut {NoStop}%
	\bibitem [{\citenamefont {\v{Z}onda}\ \emph {et~al.}(2023)\citenamefont
		{\v{Z}onda}, \citenamefont {Zalom}, \citenamefont {Novotn\'y}, \citenamefont
		{Loukeris}, \citenamefont {B\"atge},\ and\ \citenamefont
		{Pokorn\'y}}]{Zonda-2023}%
	\BibitemOpen
	\bibfield  {author} {\bibinfo {author} {\bibfnamefont {M.}~\bibnamefont
			{\v{Z}onda}}, \bibinfo {author} {\bibfnamefont {P.}~\bibnamefont {Zalom}},
		\bibinfo {author} {\bibfnamefont {T.}~\bibnamefont {Novotn\'y}}, \bibinfo
		{author} {\bibfnamefont {G.}~\bibnamefont {Loukeris}}, \bibinfo {author}
		{\bibfnamefont {J.}~\bibnamefont {B\"atge}},\ and\ \bibinfo {author}
		{\bibfnamefont {V.}~\bibnamefont {Pokorn\'y}},\ }\href
	{https://doi.org/10.1103/PhysRevB.107.115407} {\bibfield  {journal} {\bibinfo
			{journal} {Phys. Rev. B}\ }\textbf {\bibinfo {volume} {107}},\ \bibinfo
		{pages} {115407} (\bibinfo {year} {2023})}\BibitemShut {NoStop}%
	\bibitem [{\citenamefont {Pillet}\ \emph {et~al.}(2010)\citenamefont {Pillet},
		\citenamefont {Quay}, \citenamefont {Morfin}, \citenamefont {Bena},
		\citenamefont {Yeyati},\ and\ \citenamefont {Joyez}}]{Pillet-2010}%
	\BibitemOpen
	\bibfield  {author} {\bibinfo {author} {\bibfnamefont {J.-D.}\ \bibnamefont
			{Pillet}}, \bibinfo {author} {\bibfnamefont {C.~H.~L.}\ \bibnamefont {Quay}},
		\bibinfo {author} {\bibfnamefont {P.}~\bibnamefont {Morfin}}, \bibinfo
		{author} {\bibfnamefont {C.}~\bibnamefont {Bena}}, \bibinfo {author}
		{\bibfnamefont {A.~L.}\ \bibnamefont {Yeyati}},\ and\ \bibinfo {author}
		{\bibfnamefont {P.}~\bibnamefont {Joyez}},\ }\href
	{http://dx.doi.org/10.1038/nphys1811} {\bibfield  {journal} {\bibinfo
			{journal} {Nat. Phys.}\ }\textbf {\bibinfo {volume} {6}},\ \bibinfo {pages}
		{965} (\bibinfo {year} {2010})}\BibitemShut {NoStop}%
	\bibitem [{\citenamefont {Luitz}\ \emph {et~al.}(2012)\citenamefont {Luitz},
		\citenamefont {Assaad}, \citenamefont {Novotn{\'y}}, \citenamefont
		{Karrasch},\ and\ \citenamefont {Meden}}]{Luitz-2012}%
	\BibitemOpen
	\bibfield  {author} {\bibinfo {author} {\bibfnamefont {D.~J.}\ \bibnamefont
			{Luitz}}, \bibinfo {author} {\bibfnamefont {F.~F.}\ \bibnamefont {Assaad}},
		\bibinfo {author} {\bibfnamefont {T.}~\bibnamefont {Novotn{\'y}}}, \bibinfo
		{author} {\bibfnamefont {C.}~\bibnamefont {Karrasch}},\ and\ \bibinfo
		{author} {\bibfnamefont {V.}~\bibnamefont {Meden}},\ }\href
	{http://link.aps.org/doi/10.1103/PhysRevLett.108.227001} {\bibfield
		{journal} {\bibinfo  {journal} {Phys. Rev. Lett.}\ }\textbf {\bibinfo
			{volume} {108}},\ \bibinfo {pages} {227001} (\bibinfo {year}
		{2012})}\BibitemShut {NoStop}%
	\bibitem [{\citenamefont {Zalom}(2023)}]{Zalom-2023}%
	\BibitemOpen
	\bibfield  {author} {\bibinfo {author} {\bibfnamefont {P.}~\bibnamefont
			{Zalom}},\ }\href@noop {} {\bibinfo {title} {Rigorous wilsonian
			renormalization group for impurity models with a spectral gap}} (\bibinfo
	{year} {2023}),\ \Eprint {https://arxiv.org/abs/2307.07479} {arXiv:2307.07479
		[cond-mat.str-el]} \BibitemShut {NoStop}%
	\bibitem [{\citenamefont {Satori}\ \emph {et~al.}(1992)\citenamefont {Satori},
		\citenamefont {Shiba}, \citenamefont {Sakai},\ and\ \citenamefont
		{Shimizu}}]{Satori-1992}%
	\BibitemOpen
	\bibfield  {author} {\bibinfo {author} {\bibfnamefont {K.}~\bibnamefont
			{Satori}}, \bibinfo {author} {\bibfnamefont {H.}~\bibnamefont {Shiba}},
		\bibinfo {author} {\bibfnamefont {O.}~\bibnamefont {Sakai}},\ and\ \bibinfo
		{author} {\bibfnamefont {Y.}~\bibnamefont {Shimizu}},\ }\href
	{https://doi.org/10.1143/JPSJ.61.3239} {\bibfield  {journal} {\bibinfo
			{journal} {J. Phys. Soc. Japan.}\ }\textbf {\bibinfo {volume} {61}},\
		\bibinfo {pages} {3239} (\bibinfo {year} {1992})}\BibitemShut {NoStop}%
	\bibitem [{\citenamefont {Sakai}\ \emph {et~al.}(1993)\citenamefont {Sakai},
		\citenamefont {Shimizu}, \citenamefont {Shiba},\ and\ \citenamefont
		{Satori}}]{Sakai-1993}%
	\BibitemOpen
	\bibfield  {author} {\bibinfo {author} {\bibfnamefont {O.}~\bibnamefont
			{Sakai}}, \bibinfo {author} {\bibfnamefont {Y.}~\bibnamefont {Shimizu}},
		\bibinfo {author} {\bibfnamefont {H.}~\bibnamefont {Shiba}},\ and\ \bibinfo
		{author} {\bibfnamefont {K.}~\bibnamefont {Satori}},\ }\href
	{https://doi.org/10.1143/JPSJ.62.3181} {\bibfield  {journal} {\bibinfo
			{journal} {J. Phys. Soc. Japan.}\ }\textbf {\bibinfo {volume} {62}},\
		\bibinfo {pages} {3181} (\bibinfo {year} {1993})}\BibitemShut {NoStop}%
	\bibitem [{\citenamefont {Hecht}\ \emph {et~al.}(2008)\citenamefont {Hecht},
		\citenamefont {Weichselbaum}, \citenamefont {von Delft},\ and\ \citenamefont
		{Bulla}}]{Hecht-2008}%
	\BibitemOpen
	\bibfield  {author} {\bibinfo {author} {\bibfnamefont {T.}~\bibnamefont
			{Hecht}}, \bibinfo {author} {\bibfnamefont {A.}~\bibnamefont {Weichselbaum}},
		\bibinfo {author} {\bibfnamefont {J.}~\bibnamefont {von Delft}},\ and\
		\bibinfo {author} {\bibfnamefont {R.}~\bibnamefont {Bulla}},\ }\href
	{http://stacks.iop.org/0953-8984/20/i=27/a=275213} {\bibfield  {journal}
		{\bibinfo  {journal} {J. Phys.: Cond. Mat.}\ }\textbf {\bibinfo {volume}
			{20}},\ \bibinfo {pages} {275213} (\bibinfo {year} {2008})}\BibitemShut
	{NoStop}%
	\bibitem [{\citenamefont {Siano}\ and\ \citenamefont
		{Egger}(2004)}]{Siano-2004}%
	\BibitemOpen
	\bibfield  {author} {\bibinfo {author} {\bibfnamefont {F.}~\bibnamefont
			{Siano}}\ and\ \bibinfo {author} {\bibfnamefont {R.}~\bibnamefont {Egger}},\
	}\href {http://link.aps.org/doi/10.1103/PhysRevLett.93.047002} {\bibfield
		{journal} {\bibinfo  {journal} {Phys. Rev. Lett.}\ }\textbf {\bibinfo
			{volume} {93}},\ \bibinfo {pages} {047002} (\bibinfo {year}
		{2004})}\BibitemShut {NoStop}%
	\bibitem [{\citenamefont {Siano}\ and\ \citenamefont
		{Egger}(2005)}]{Siano-2005rep}%
	\BibitemOpen
	\bibfield  {author} {\bibinfo {author} {\bibfnamefont {F.}~\bibnamefont
			{Siano}}\ and\ \bibinfo {author} {\bibfnamefont {R.}~\bibnamefont {Egger}},\
	}\href {http://link.aps.org/doi/10.1103/PhysRevLett.94.229702} {\bibfield
		{journal} {\bibinfo  {journal} {Phys. Rev. Lett.}\ }\textbf {\bibinfo
			{volume} {94}},\ \bibinfo {pages} {229702} (\bibinfo {year}
		{2005})}\BibitemShut {NoStop}%
	\bibitem [{\citenamefont {Gubernatis}(2016)}]{Gubernatis-2016}%
	\BibitemOpen
	\bibfield  {author} {\bibinfo {author} {\bibfnamefont {J.~E.}\ \bibnamefont
			{Gubernatis}},\ }\href@noop {} {\emph {\bibinfo {title} {Quantum Monte Carlo
				methods : algorithms for lattice models}}}\ (\bibinfo  {publisher} {Cambridge
		University Press},\ \bibinfo {address} {New York, NY},\ \bibinfo {year}
	{2016})\BibitemShut {NoStop}%
	\bibitem [{\citenamefont {Karrasch}\ \emph {et~al.}(2008)\citenamefont
		{Karrasch}, \citenamefont {Oguri},\ and\ \citenamefont
		{Meden}}]{Karrasch-2008}%
	\BibitemOpen
	\bibfield  {author} {\bibinfo {author} {\bibfnamefont {C.}~\bibnamefont
			{Karrasch}}, \bibinfo {author} {\bibfnamefont {A.}~\bibnamefont {Oguri}},\
		and\ \bibinfo {author} {\bibfnamefont {V.}~\bibnamefont {Meden}},\ }\href
	{http://link.aps.org/doi/10.1103/PhysRevB.77.024517} {\bibfield  {journal}
		{\bibinfo  {journal} {Phys. Rev. B}\ }\textbf {\bibinfo {volume} {77}},\
		\bibinfo {pages} {024517} (\bibinfo {year} {2008})}\BibitemShut {NoStop}%
	\bibitem [{\citenamefont {Pokorn\'y}\ and\ \citenamefont
		{\v{Z}onda}(2023)}]{Zonda-2022}%
	\BibitemOpen
	\bibfield  {author} {\bibinfo {author} {\bibfnamefont {V.}~\bibnamefont
			{Pokorn\'y}}\ and\ \bibinfo {author} {\bibfnamefont {M.}~\bibnamefont
			{\v{Z}onda}},\ }\href {https://doi.org/10.1103/PhysRevB.107.155111}
	{\bibfield  {journal} {\bibinfo  {journal} {Phys. Rev. B}\ }\textbf {\bibinfo
			{volume} {107}},\ \bibinfo {pages} {155111} (\bibinfo {year}
		{2023})}\BibitemShut {NoStop}%
	\bibitem [{\citenamefont {Baran}\ \emph {et~al.}(2023)\citenamefont {Baran},
		\citenamefont {Frost},\ and\ \citenamefont {Paaske}}]{Paaske-arxiv}%
	\BibitemOpen
	\bibfield  {author} {\bibinfo {author} {\bibfnamefont {V.~V.}\ \bibnamefont
			{Baran}}, \bibinfo {author} {\bibfnamefont {E.~J.~P.}\ \bibnamefont
			{Frost}},\ and\ \bibinfo {author} {\bibfnamefont {J.}~\bibnamefont
			{Paaske}},\ }\Eprint {https://arxiv.org/abs/2307.11646} {arXiv:2307.11646
		[cond-mat.supr-con]}  (\bibinfo {year} {2023})\BibitemShut {NoStop}%
	\bibitem [{\citenamefont {Bargerbos}\ \emph {et~al.}(2022)\citenamefont
		{Bargerbos}, \citenamefont {Pita-Vidal}, \citenamefont
		{\ifmmode~\check{Z}\else \v{Z}\fi{}itko}, \citenamefont {\'Avila},
		\citenamefont {Splitthoff}, \citenamefont {Gr\"unhaupt}, \citenamefont
		{Wesdorp}, \citenamefont {Andersen}, \citenamefont {Liu}, \citenamefont
		{Kouwenhoven}, \citenamefont {Aguado}, \citenamefont {Kou},\ and\
		\citenamefont {van Heck}}]{Bargerbos-2022}%
	\BibitemOpen
	\bibfield  {author} {\bibinfo {author} {\bibfnamefont {A.}~\bibnamefont
			{Bargerbos}}, \bibinfo {author} {\bibfnamefont {M.}~\bibnamefont
			{Pita-Vidal}}, \bibinfo {author} {\bibfnamefont {R.}~\bibnamefont
			{\ifmmode~\check{Z}\else \v{Z}\fi{}itko}}, \bibinfo {author} {\bibfnamefont
			{J.}~\bibnamefont {\'Avila}}, \bibinfo {author} {\bibfnamefont {L.~J.}\
			\bibnamefont {Splitthoff}}, \bibinfo {author} {\bibfnamefont
			{L.}~\bibnamefont {Gr\"unhaupt}}, \bibinfo {author} {\bibfnamefont {J.~J.}\
			\bibnamefont {Wesdorp}}, \bibinfo {author} {\bibfnamefont {C.~K.}\
			\bibnamefont {Andersen}}, \bibinfo {author} {\bibfnamefont {Y.}~\bibnamefont
			{Liu}}, \bibinfo {author} {\bibfnamefont {L.~P.}\ \bibnamefont
			{Kouwenhoven}}, \bibinfo {author} {\bibfnamefont {R.}~\bibnamefont {Aguado}},
		\bibinfo {author} {\bibfnamefont {A.}~\bibnamefont {Kou}},\ and\ \bibinfo
		{author} {\bibfnamefont {B.}~\bibnamefont {van Heck}},\ }\href
	{https://doi.org/10.1103/PRXQuantum.3.030311} {\bibfield  {journal} {\bibinfo
			{journal} {PRX Quantum}\ }\textbf {\bibinfo {volume} {3}},\ \bibinfo {pages}
		{030311} (\bibinfo {year} {2022})}\BibitemShut {NoStop}%
	\bibitem [{\citenamefont {Pavešič}\ \emph {et~al.}(2023)\citenamefont
		{Pavešič}, \citenamefont {Aguado},\ and\ \citenamefont
		{Žitko}}]{Zitko-arxiv}%
	\BibitemOpen
	\bibfield  {author} {\bibinfo {author} {\bibfnamefont {L.}~\bibnamefont
			{Pavešič}}, \bibinfo {author} {\bibfnamefont {R.}~\bibnamefont {Aguado}},\
		and\ \bibinfo {author} {\bibfnamefont {R.}~\bibnamefont {Žitko}},\ }\Eprint
	{https://arxiv.org/abs/2304.12456} {arXiv:2304.12456 [cond-mat.mes-hall]}
	(\bibinfo {year} {2023})\BibitemShut {NoStop}%
	\bibitem [{Note1()}]{Note1}%
	\BibitemOpen
	\bibinfo {note} {The mechanism behind both effects is novel and relies on the
		full phase control over the three terminal architecture of the device and the
		presence of Coulomb interactions in the junction region.}\BibitemShut {Stop}%
	\bibitem [{\citenamefont {Zalom}\ \emph {et~al.}(2021)\citenamefont {Zalom},
		\citenamefont {Pokorn\'y},\ and\ \citenamefont {Novotn\'y}}]{Zalom-2021}%
	\BibitemOpen
	\bibfield  {author} {\bibinfo {author} {\bibfnamefont {P.}~\bibnamefont
			{Zalom}}, \bibinfo {author} {\bibfnamefont {V.}~\bibnamefont {Pokorn\'y}},\
		and\ \bibinfo {author} {\bibfnamefont {T.}~\bibnamefont {Novotn\'y}},\ }\href
	{https://doi.org/10.1103/PhysRevB.103.035419} {\bibfield  {journal} {\bibinfo
			{journal} {Phys. Rev. B}\ }\textbf {\bibinfo {volume} {103}},\ \bibinfo
		{pages} {035419} (\bibinfo {year} {2021})}\BibitemShut {NoStop}%
	\bibitem [{sup()}]{supp}%
	\BibitemOpen
	\href@noop {} {}\bibinfo {howpublished}
	{\url{URL_will_be_inserted_by_publisher}}\BibitemShut {NoStop}%
	\bibitem [{Note2()}]{Note2}%
	\BibitemOpen
	\bibinfo {note} {The dot Green function actually changes, but it has no
		physical consequences, for a detailed account of this issue for the
		two-terminal setup see Ref.~\cite {Kadlecova-2017}, the situation for
		multi-terminal setup is completely analogous.}\BibitemShut {Stop}%
	\bibitem [{\citenamefont {Kadlecov{\'a}}\ \emph {et~al.}(2017)\citenamefont
		{Kadlecov{\'a}}, \citenamefont {{\v Z}onda},\ and\ \citenamefont
		{Novotn{\'y}}}]{Kadlecova-2017}%
	\BibitemOpen
	\bibfield  {author} {\bibinfo {author} {\bibfnamefont {A.}~\bibnamefont
			{Kadlecov{\'a}}}, \bibinfo {author} {\bibfnamefont {M.}~\bibnamefont {{\v
					Z}onda}},\ and\ \bibinfo {author} {\bibfnamefont {T.}~\bibnamefont
			{Novotn{\'y}}},\ }\href {https://doi.org/10.1103/PhysRevB.95.195114}
	{\bibfield  {journal} {\bibinfo  {journal} {Phys. Rev. B}\ }\textbf {\bibinfo
			{volume} {95}},\ \bibinfo {pages} {195114} (\bibinfo {year}
		{2017})}\BibitemShut {NoStop}%
	\bibitem [{Note3()}]{Note3}%
	\BibitemOpen
	\bibinfo {note} {The Hamiltonian is invariant under $d^{\protect \vphantom
			{\dagger }}_{\uparrow }\rightarrow d^{\protect \vphantom {\dagger
		}}_{\downarrow }$, $d^{\protect \vphantom {\dagger }}_{\downarrow
		}\rightarrow d^{\protect \vphantom {\dagger }}_{\uparrow }$ followed by a
		complex conjugation. Note that this antiunitary symmetry is different from
		time-reversal symmetry attained for two-terminal systems at $\varphi =0$ as
		noted in Ref.~\cite {Zitko-arxiv}.}\BibitemShut {Stop}%
	\bibitem [{Note4()}]{Note4}%
	\BibitemOpen
	\bibinfo {note} {The first BZ collapses to $1D$ in {\protect \bf C} and the
		green line thus corresponds always to one and the same $\chi =0$
		point.}\BibitemShut {Stop}%
	\bibitem [{Bau(2007)}]{Bauer-2007}%
	\BibitemOpen
	\href {http://stacks.iop.org/0953-8984/19/i=48/a=486211} {\bibfield
		{journal} {\bibinfo  {journal} {J. Phys.: Cond. Mat.}\ }\textbf {\bibinfo
			{volume} {19}},\ \bibinfo {pages} {486211} (\bibinfo {year}
		{2007})}\BibitemShut {NoStop}%
	\bibitem [{\citenamefont {Zalom}\ and\ \citenamefont
		{Novotn\'{y}}(2021)}]{Zalom-2021r}%
	\BibitemOpen
	\bibfield  {author} {\bibinfo {author} {\bibfnamefont {P.}~\bibnamefont
			{Zalom}}\ and\ \bibinfo {author} {\bibfnamefont {T.}~\bibnamefont
			{Novotn\'{y}}},\ }\href {https://doi.org/10.1103/PhysRevB.104.035437}
	{\bibfield  {journal} {\bibinfo  {journal} {Phys. Rev. B}\ }\textbf {\bibinfo
			{volume} {104}},\ \bibinfo {pages} {035437} (\bibinfo {year}
		{2021})}\BibitemShut {NoStop}%
	\bibitem [{\citenamefont {Zalom}\ and\ \citenamefont {\ifmmode~\check{Z}\else
			\v{Z}\fi{}onda}(2022)}]{Zalom-2022}%
	\BibitemOpen
	\bibfield  {author} {\bibinfo {author} {\bibfnamefont {P.}~\bibnamefont
			{Zalom}}\ and\ \bibinfo {author} {\bibfnamefont {M.}~\bibnamefont
			{\ifmmode~\check{Z}\else \v{Z}\fi{}onda}},\ }\href
	{https://doi.org/10.1103/PhysRevB.105.205412} {\bibfield  {journal} {\bibinfo
			{journal} {Phys. Rev. B}\ }\textbf {\bibinfo {volume} {105}},\ \bibinfo
		{pages} {205412} (\bibinfo {year} {2022})}\BibitemShut {NoStop}%
	\bibitem [{\citenamefont {Estrada Salda\~na}\ \emph {et~al.}(2018)\citenamefont
		{Estrada Salda\~na}, \citenamefont {Vekris}, \citenamefont {Steffensen},
		\citenamefont {\ifmmode~\check{Z}\else \v{Z}\fi{}itko}, \citenamefont
		{Krogstrup}, \citenamefont {Paaske}, \citenamefont {Grove-Rasmussen},\ and\
		\citenamefont {Nyg\aa{}rd}}]{Saldana-2018}%
	\BibitemOpen
	\bibfield  {author} {\bibinfo {author} {\bibfnamefont {J.~C.}\ \bibnamefont
			{Estrada Salda\~na}}, \bibinfo {author} {\bibfnamefont {A.}~\bibnamefont
			{Vekris}}, \bibinfo {author} {\bibfnamefont {G.}~\bibnamefont {Steffensen}},
		\bibinfo {author} {\bibfnamefont {R.}~\bibnamefont {\ifmmode~\check{Z}\else
				\v{Z}\fi{}itko}}, \bibinfo {author} {\bibfnamefont {P.}~\bibnamefont
			{Krogstrup}}, \bibinfo {author} {\bibfnamefont {J.}~\bibnamefont {Paaske}},
		\bibinfo {author} {\bibfnamefont {K.}~\bibnamefont {Grove-Rasmussen}},\ and\
		\bibinfo {author} {\bibfnamefont {J.}~\bibnamefont {Nyg\aa{}rd}},\ }\href
	{https://doi.org/10.1103/PhysRevLett.121.257701} {\bibfield  {journal}
		{\bibinfo  {journal} {Phys. Rev. Lett.}\ }\textbf {\bibinfo {volume} {121}},\
		\bibinfo {pages} {257701} (\bibinfo {year} {2018})}\BibitemShut {NoStop}%
	\bibitem [{\citenamefont {\v{Z}itko}(2021)}]{NRGzenodo}%
	\BibitemOpen
	\bibfield  {author} {\bibinfo {author} {\bibfnamefont {R.}~\bibnamefont
			{\v{Z}itko}},\ }\Eprint {https://arxiv.org/abs/NRG Ljubljana (8f90ac4),
		Zenodo, https://doi.org/10.5281/zenodo.4841076} {NRG Ljubljana (8f90ac4),
		Zenodo, https://doi.org/10.5281/zenodo.4841076}  (\bibinfo {year}
	{2021})\BibitemShut {NoStop}%
\end{thebibliography}
\end{document}


\title{{Supplementary Material: Hidden symmetry in multi-terminal Josephson junctions}}
	
	\author{Peter Zalom}
	\email{zalomp@fzu.cz}
	\affiliation{Institute of Physics, Czech Academy of Sciences, Na Slovance 2, CZ-18200 Praha 8, Czech Republic}
	
	\author{M. \v{Z}onda}
	\email{martin.zonda@matfyz.cuni.cz}
	
	\affiliation{Department of Condensed Matter Physics, Faculty of Mathematics and Physics, Charles University, Ke Karlovu 5, CZ-121 16 Praha 2, Czech Republic}
	
	\author{T. Novotn\'y}
	\email{tno@karlov.mff.cuni.cz}
	\affiliation{Department of Condensed Matter Physics, Faculty of Mathematics and Physics, Charles University, Ke Karlovu 5, CZ-121 16 Praha 2, Czech Republic}

	\date{\today}

	\maketitle
	\section{Model}
	For convenience we repeat here the Supercconducting Anderson Impurity model (SC-AIM) used in the main text for the description of a general multi-terminal setup. Its Hamiltonian reads
	\begin{subequations}
		\begin{align}
		\hat{H}&=\hat{H}_d+\sum_{j=1}^{n}\left(\hat{H}_{j,SC}+\hat{H}_{j,T}\right)
		\end{align}
		with  $j \in \{1,\ldots n \}$ denoting a given lead with a superconducting phase $\varphi_j$ and
		\begin{align}
		\hat{H}_d
		&=
		\sum_{\sigma} 
		\varepsilon_{d}
		d^{\dagger}_{\sigma}
		d^{\vphantom{\dagger}}_{ \sigma}
		+
		U
		d^{\dagger}_{\uparrow}
		d^{\vphantom{\dagger}}_{ \uparrow}
		d^{\dagger}_{\downarrow}
		d^{\vphantom{\dagger}}_{ \downarrow},
		\label{eq:dotH}
		\\
		\hat{H}_{j,SC}
		&=
		\sum_{\mathbf{k}\sigma} 
		\, \varepsilon_{\mathbf{k}j}
		c^{\dagger}_{\mathbf{k} j\sigma}
		c^{\vphantom{\dagger}}_{\mathbf{k} j\sigma}
		-\sum_{\mathbf{k}}
		\left(
		\Delta_{j}
		c^{\dagger}_{\mathbf{k}j \uparrow} 
		c^{\dagger}_{-\mathbf{k}j \downarrow}
		+
		\textit{H.c.}\right),\label{eq:bcsH}
		\\
		\hat{H}_{j,T} 
		&=
		\sum_{\mathbf{k} \sigma} \,
		\left(V^*_{\mathbf{k}j}  
		c^{\dagger}_{\mathbf{k}j\sigma}
		d^{\vphantom{\dagger}}_{\sigma} 
		+ 
		V_{\mathbf{k}j}  
		d^{\dagger}_{\sigma}
		c^{\vphantom{\dagger}}_{\mathbf{k}j\sigma}\right),
		\label{eq:tunnelH}
		\end{align}
	\end{subequations}
	where $c^{\dagger}_{\mathbf{k} j\sigma}$ ($c^{\vphantom{\dagger}}_{\mathbf{k}j\sigma}$) creates (annihilates) an electron of spin $\sigma \in \{\uparrow, \downarrow \}$,  quasi-momentum $\mathbf{k}$, and energy $\varepsilon_{\mathbf{k}j}$ in the lead $j$, while $d^{\dagger}_{\sigma}$ ($d^{\vphantom{\dagger}}_{\sigma}$) creates (annihilates) a dot electron of spin $\sigma$. The first term \eqref{eq:dotH} describes the quantum dot with a single energy level $\varepsilon_{d}$, where electrons of opposite spins repel each other by an effective Coulomb interaction $U$. In our paper, we assume a particle-hole symmetric case fixed by $\varepsilon_d=-U/2$ for simplicity. Nevertheless, the geometric factor $\chi$ introduced in the main text does not depend on $\varepsilon_d$. Consequently, the presented methods can be extended beyond such a constraint. The second term~\eqref{eq:bcsH} describes the superconducting leads. Here, we assume the same gap $\Delta$, and therefore $\Delta_{j} \equiv \Delta e^{i\varphi_{j}}$.  We also assume the same bandwidth $2B$ and dispersions in all leads, which reflects the typical experimental situations, where all terminals are made from the same material. The last term~\eqref{eq:tunnelH} is the hybridization between the leads and the dot. The coupling to the leads is characterized by the tunneling strengths ($\hbar=1$) $\Gamma_j\equiv\pi\sum_{\mathbf{k}}|	V_{\mathbf{k}j}|^2\delta(\omega-\varepsilon_{\mathbf{k}j})$ that are assumed to be energy independent. For convenience, we also introduce the total tunneling strength $\Gamma\equiv\sum_{j=1}^{n}\Gamma_j$ and the relative couplings $\gamma_j\equiv\Gamma_j/\Gamma$, so $\sum_{j=1}^{n}\gamma_j=1$. 
	
	\section{Tunneling density of states for superconducting Anderson models}
	
	Using the Nambu spinor $D^{\dagger} =  \left(d^{\dagger}_{\uparrow}, 
	d^{\vphantom{\dagger}}_{\downarrow} \right)$, we can extract the tunneling self-energy $\mathbb{\Sigma}$ due to the superconducting leads by the equation of motion technique described in detail in Refs.~\cite{Zalom-2021}. For $n$ superconducting leads, we obtain
	\begin{eqnarray}
	\mathbb{\Sigma}(z)
	=
	\sum_{\mathbf{k}}
	\sum_{j=1}^n
	\mathbb{V}_{j\mathbf{k}}
	\left(
	z \cdot \mathbb{1} - \mathbb{E}_{j\mathbf{k}}
	\right)^{-1}
	\mathbb{V}_{j\mathbf{k}}
	\label{sup:sigma}
	\end{eqnarray}
	with $z$ being an arbitrary complex number. Later we set it to $z=\omega^+\equiv \omega+i\eta $ with $\omega$ being a real frequency and $\eta$ being an infinitesimally small positive number. Additionally,
	\begin{align}
	\mathbb{E}_{j\mathbf{k}}  
	&=
	-\Delta_j C_{j} \sigma_x
	+\Delta_j S_{j} \sigma_y
	+ \varepsilon_{\mathbf{k}j} \sigma_z,
	\label{Ealpha}
	\\
	\mathbb{V}_{j\mathbf{k}}
	&=
	V_{j\mathbf{k}}  \, \sigma_z,
	\label{Valpha}
	\end{align}
	where $\sigma_i$ are the Pauli matrices with $i \in x,y,z$, $C_j\equiv \cos\varphi_j$, $S_j\equiv \sin\varphi_j$ and $V_{j\mathbf{k}}$ are the hybridizations for the individual leads $j$. The inverse matrix appearing in Eq.~(\ref{sup:sigma}) is evaluated using the identity $(u \mathbb{1}+\vec{v}\cdot\vec{\sigma})^{-1}=(u \mathbb{1}-\vec{v}\cdot\vec{\sigma})/(u^2-\vec{v}\cdot\vec{v})$. It yields
	\begin{equation}
	\left(
	\omega^+ \mathbb{1} - \mathbb{E}_{j\mathbf{k}}
	\right)^{-1}
	= \frac{
		\omega \mathbb{1} 
		- \Delta C_{j} \sigma_x
		+ \Delta S_{j} \sigma_y
		+\varepsilon_{\mathbf{k}j} \sigma_z
	}{(\omega+i\eta)^2 - \Delta^2 - \varepsilon_{\mathbf{k}j}^2}.
	\end{equation}
	Furthermore, since $\sigma_z(u \mathbb{1}+v_x\sigma_x+v_y\sigma_y+v_z\sigma_z)\sigma_z=u \mathbb{1}-v_x\sigma_x-v_y\sigma_y+v_z\sigma_z$, we get
	\begin{equation}
	\mathbb{V}_{j\mathbf{k}}
	\left(
	\omega^+ \mathbb{1} - \mathbb{E}_{j\mathbf{k}}
	\right)^{-1}
	\mathbb{V}_{j\mathbf{k}}=
	\hspace{3cm}
	\nonumber
	\\
	V_{j\mathbf{k}}^2
	\frac{
		\omega \mathbb{1} 
		+ \Delta C_{j} \sigma_x
		- \Delta S_{j} \sigma_y
		+\varepsilon_{\mathbf{k}j} \sigma_z
	}
	{\omega^2 - \Delta^2 - \varepsilon_{\mathbf{k}j}^2 + i\eta\, \mathrm{sgn}(\omega)
	}.
	\label{B6}
	\end{equation}
	Assuming constant and $\bf k$-independent tunneling DOS and the same bandwidth $2B$ in all superconducting leads, we obtain
	\begin{equation}
	\mathbb{\Sigma}(\omega^+)
	=
	\sum_{j=1}^n 
	\left[
	\omega \Gamma_j \mathbb{1} 
	+ \Delta \Gamma_j C_{j} \sigma_x
	- \Delta \Gamma_jS_{j} \sigma_y
	\right]
	F(\omega^+),
	\end{equation}
	where the term proportional to $\sigma_z$ vanished due to the integrand being an odd function of $\varepsilon$ while
	\begin{equation}
	F(\omega^+)
	\equiv
	\frac{1}{\pi}
	\int_{-B}^{B}
	\frac{d\varepsilon}{\omega^2 - \Delta^2 - \varepsilon^2 + i\eta\, \mathrm{sgn}(\omega)}\\
	=\frac{1}{\pi\sqrt{(\omega+ i\eta)^2 - \Delta^2 }}\ln\frac{\sqrt{(\omega+ i\eta)^2 - \Delta^2}+B}{\sqrt{(\omega+ i\eta)^2 - \Delta^2}-B}.
	\end{equation}
	Taking the $\eta \rightarrow 0$ limit, we arrive at 
	\begin{equation}
	F(\omega^+)
	=
	\begin{cases}
	-\frac{2}{\pi\sqrt{\Delta^2-\omega^2}} \arctan
	\left(\frac{B}{\sqrt{\Delta^2-\omega^2}}\right),
	&
	\text{for }|\omega|<\Delta
	\\
	-\frac{i\, \mathrm{sgn}(\omega)}{\sqrt{\omega^2-\Delta^2}}
	+
	\frac{\ln \left( \frac{B+\sqrt{\omega^2-\Delta^2}}{B-\sqrt{\omega^2-\Delta^2}}\right)}{\pi\sqrt{\omega^2-\Delta^2}},
	&
	\text{for }\Delta<|\omega|<B,
	\end{cases}
	\end{equation}
	which is apparently a universal function for all leads as stated in the main text.
	
	The resulting $\mathbb{\Sigma}(\omega^+)$ thus contains an universal prefactor dependent on the geometry and particular hybridizations of the superconducting system multiplied by a universal factor of $F(\omega^+)$, which has a non-zero imaginary part only outside of the gap region. All effects of the finite-sized band appear in its real part, which is non-zero in the whole band. In the limit $B\rightarrow\infty$, the real part vanishes outside the gap. 
	
	Let us now turn our attention to the structure of the prefactor for two special cases with $n=2$ and $n=3$. For the former one it is customary to use the gauge of a symmetric phase drop with $\varphi_L=-\varphi_R=\varphi/2$, which yields
	\begin{equation}
	\mathbb{\Sigma}_{n=2}(\omega^+)
	=
	\Gamma_S
	\left[	
	\omega \mathbb{1} 
	+ \Delta \cos\left(\frac{\varphi}{2}\right) \sigma_x
	\right]
	F(\omega^+).
	\label{eq:two_channel}
	\end{equation}
	The results was already derived multiple times, see for example Refs.~\cite{Bauer-2007,Zalom-2021,Zalom-2021r,Zalom-2022}. We note that the formerly used gauge is actually just a special case of the gauge shift by $\delta$ as proposed in the main text which makes $\bf \chi$ purely real. For $n \geq 3$, such a gauge leads to 
	\begin{equation}
	\mathbb{\Sigma}_{n=3}(\omega^+)
	=
	\Gamma_S
	\left[	
	\omega \mathbb{1} 
	+ \Delta \chi \sigma_x
	\right]
	F(\omega^+)
	\end{equation}
	with a geometrical factor $\chi$.

	\section{Gauge independence of the geometrical factor $\bf \chi$}
	
	While the expression $\chi= |{\bf \chi}| = |\sum_{j=1}^n \gamma_{j} e^{-i\varphi_j}|$ can be applied in a straightforward way, it is now our intention to demonstrate a different conceptual approach. We first apply a gauge shifting phase $\delta$ to rotate $\bf \chi$ so it points in the direction of real axis afterwards. Thus, we demand
	\begin{align}
	\chi
	&=
	\sum_{j}
	\gamma_{j} \cos \left( \varphi_j -\delta  \right),
	\\
	0
	&=
	\sum_{j}
	\gamma_{j} \sin \left( \varphi_j -\delta  \right),
	\\
	\end{align}
	which yields
	\begin{equation}
	\tan{\delta}
	=
	\frac
	{\sum_{j=1}^n \gamma_{j} \sin \varphi_{j}}
	{\sum_{j=1}^n \gamma_{j} \cos \varphi_{j}}.
	\label{eq:gauge}
	\end{equation}
	Using trigonometric relations $\cos\delta=1/\sqrt{1+\tan^2\delta}$ and $\sin\delta=\tan\delta/\sqrt{1+\tan^2\delta}$ we then obtain
	\begin{align}
	\chi
	&=
	\sum_{i=1}^n
	\gamma_i\cos\varphi_i\cos\delta
	+
	\gamma_i\sin\varphi_i\sin\delta
	=
	\sum_{i=1}^n
	\gamma_i
	\frac
	{
		\cos\varphi_i + \sin\varphi_i
		\frac
		{
			\sum_{j=1}^n
			\gamma_j\sin\varphi_j
		}
		{	
			\sum_{j=1}^n
			\gamma_j\cos\varphi_j
		}
	}
	{
		\sqrt{
			\frac
			{
				\left(\sum_{j=1}^n
				\gamma_j\cos\varphi_j\right)^2
				+
				\left(\sum_{j=1}^n
				\gamma_j\sin\varphi_j\right)^2
			}
			{	
				\left(\sum_{j=1}^n
				\gamma_j\cos\varphi_i\right)^2
			}
		}
	}=
	\nonumber
	\\
	&=
	\sum_{i=1}^n
	\gamma_i
	\frac
	{
		\cos\varphi_i 
		\left(\sum_{j=1}^n
		\gamma_j\cos\varphi_j\right)
		+ 
		\sin\varphi_i
		\left(\sum_{j=1}^n
		\gamma_j\sin\varphi_j\right)
	}
	{
		\sqrt{
			\left(\sum_{j=1}^n \gamma_j\cos\varphi_j\right)^2
			+
			\left(\sum_{j=1}^n \gamma_j\sin\varphi_j\right)^2
		}
	}
	=
	\sqrt
	{
		\left(\sum_{i=1}^n
		\gamma_i\cos\varphi_i\right)^2
		+
		\left(\sum_{i=1}^n
		\gamma_i\sin\varphi_i\right)^2
	}=
	\nonumber
	\\
	&=
	\sqrt{\sum_{i=1}^n
		\sum_{j=1}^n
		\left(
		\gamma_i\gamma_j\cos\varphi_i\cos\varphi_j
		+
		\gamma_i\gamma_j\sin\varphi_i\sin\varphi_j
		\right)}
	=
	\sqrt{\sum_{i=1}^n \gamma_i^2
		+
		\sum_{i>j}
		\gamma_i\gamma_j
		\left(
		\cos\varphi_i\cos\varphi_j
		+
		\sin\varphi_i\sin\varphi_j
		\right)}=
	\nonumber
	\\
	&=
	\sqrt
	{
		1
		-
		4\sum_{i>j}
		\gamma_i\gamma_j
		\sin^2
		\left(
		\frac{\varphi_i-\varphi_j}{2}
		\right).
	}
	\label{eq:chi_srt2}
	\end{align}
	As required, $\chi$ is manifestly gauge invariant and depends only on phase differences $\varphi_j-\varphi_l$.
	
	\section{High-symmetry point $\chi=0$}
	
	Although the conditions for the existence of the high-symmetry points $\chi=0$ in a multi-terminal setup can be derived directly from the expression for the gauge-invariant magnitude of the geometric factor in Eq.~\eqref{eq:chi_srt2}, let us take the advantage of working with the complex-valued $\bm{\chi}$. 
	
	\begin{figure}[ht]
		\includegraphics[width=0.8\columnwidth]{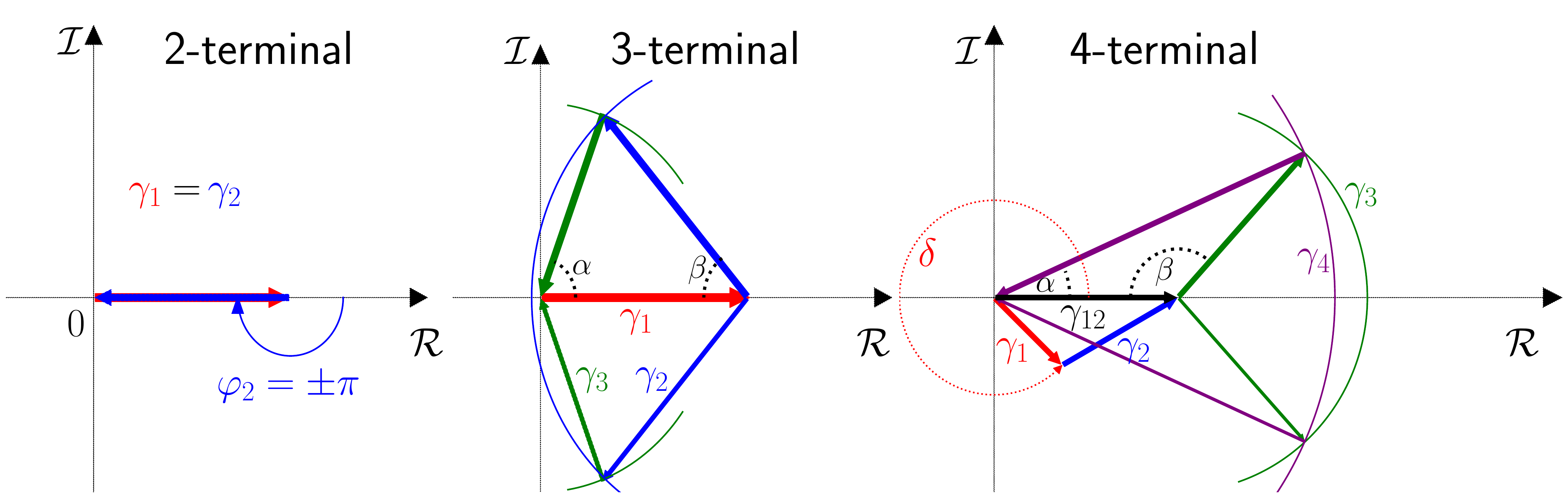}
		\caption{Graphical solutions for the high-symmetry point $\chi=0$ for two-, three- and four-terminal setup. Two- and three-terminal setups are rotated to have $\varphi_1=0$. The four-terminal setup is rotated by $\delta$ for $\bm{\gamma}_{12}=\gamma_1 e^{-i \varphi_1} + \gamma_2 e^{-i \varphi_2}$ to be real. 
			\label{fig:chi0_vectors}}
	\end{figure} 
	
	To obtain $\bm{\chi}=0$, a closed loop, which begins and ends at zero (see Fig.~\ref{fig:chi0_vectors}), must be formed by the complex vectors $\gamma_j e^{i \varphi_j}$, which satisfy $\sum_j\gamma_j = 1$. Using gauge invariance, we set $\varphi_1 \equiv 0$. Then, any general solution $\bm{\chi}=0$ has to fulfill the equations
	\begin{align}
	1 &= \sum_{j>1} \gamma_{j}\left[1-\cos \left( \varphi_j\right)\right], \nonumber
	\\
	0 &= \sum_{j>1} \gamma_{j} \sin \left( \varphi_j\right).
	\label{eq:chi0seq}
	\end{align}
	It is expected, that in the experiments the relative couplings $\gamma_j$ are fixed but phases $\varphi_j$ can be tuned. Therefore, we seek solutions for $\varphi_j$. For two- and three-terminal setups, this can be easily done and understood by following the graphical representation in Fig.~\ref{fig:chi0_vectors}. Clearly, for $n=2$ the solution exists only for the symmetric coupling $\gamma_1=\gamma_2=0.5$ and reads $\varphi_2=\pm\pi$. For $n=3$ the vectors have to form a triangle, therefore, the solution exists only when $\gamma_i+\gamma_j \geq \gamma_k$ for all permutations of the indices $\{1,2,3\}$. This reduces to the condition $\gamma_{\text{max}}\leq1/2$, where $\gamma_{\text{max}}=\text{max}(\{\gamma_j\})$, because $\gamma_1 = 1 -\gamma_2 - \gamma_3$. If the condition is satisfied, the solutions can be easily found by finding the intersections of two circles. The first is centered at zero and has the radius $\gamma_3$, while the second possesses its center at $(\gamma_1,0)$ and has the radius $\gamma_2$. All we need to calculate are thus the angles $\alpha$ and $\beta$ denoted in accord with Fig.~\ref{fig:chi0_vectors}:
	\begin{equation}
	\alpha = \arccos \frac{\gamma^2_1 - \gamma^2_2 + \gamma^2_3}{2\gamma_1\gamma_3},\,\,\,\,  
	\beta = \arccos \frac{\gamma^2_1 + \gamma^2_2 - \gamma^2_3}{2\gamma_1\gamma_2}.
	\end{equation}
	Due to the intersections being placed symmetrically in the upper and lower quadrants, and the additional $2\pi$ periodicity of the trigonometric functions, the following $\varphi_2$ and $\varphi_3$ solve for $\bm{\chi}$:
	\begin{align}
	\varphi_2 &= \pi \mp \arccos \frac{1-2(\gamma_2 +\gamma_3 -\gamma_2\gamma_3) +2\gamma^2_2}{2(1-\gamma_2-\gamma_3)\gamma_2} + 2\pi z_2, \\
	\varphi_3 &= \pi \pm \arccos  \frac{1-2(\gamma_2 +\gamma_3 -\gamma_2\gamma_3) +2 \gamma^2_3}{2(1-\gamma_2-\gamma_3)\gamma_3} + 2\pi z_3
	\end{align}
	with $z_j\in \mathbb{Z}$. 
	
	The solution becomes rather complex for more than three terminals. This is because Eq.~\eqref{eq:chi0seq} is overdetermined, that is, we have more unknown quantities than equations. However, the problem can be solved iteratively if we treat all but two phases as fixed parameters. As an example, let us show the solution for the four-terminal setup. For convenience, we denote the relative couplings in ascending order according to their strengths $\gamma_1\leq\gamma_2\leq\gamma_3\leq\gamma_4$. Then, the necessary and sufficient condition for the $\bm{\chi}=0$ solution to exist is $\sum_{j=1}^3\gamma_j \geq \gamma_4$. Because $\gamma_1=1-\sum_{j=2}^4\gamma_j$, we get $\gamma_4\leq 1/2$ or simply $\gamma_{\text{max}} \leq 1/2$ for any general multi-terminal case.
	
	Now, we take $\varphi_2$ as a fixed parameter. In the first step, we construct an auxiliary complex vector (complex number) $\bm{\gamma}_{12}=\gamma_1 + \gamma_2 e^{-i \varphi_2}$, where again, we set $\varphi_1\equiv0$. We rotate the loop around zero by the angle $\delta$ so that $\bm{\gamma}_{12}$ aligns with the $\mathcal{R}$ axis (see Fig.~\ref{fig:chi0_vectors}). This can be done using Eq.~\eqref{eq:gauge} with $n$ set to $2$. The magnitude of $\gamma_{12}\equiv\|\bm{\gamma}_{12}\|$ is then however
	\begin{equation}
	\gamma_{12} = \sqrt{\left(\gamma_1+\gamma_2\right)^2-4\gamma_1\gamma_2\sin^2\left(\frac{\varphi_2}{2}\right)}
	\end{equation}
	because $\gamma_1+\gamma_2 \neq 1$. From this result, we continue following the same steps as for the three-terminal setup and determine the angles
	\begin{equation}
	\alpha = \arccos \frac{\gamma^2_{12} - \gamma^2_3 + \gamma^2_4}{2\gamma_{12}\gamma_4},\,\,\,\,  
	\beta = \arccos \frac{\gamma^2_{12} + \gamma^2_3 - \gamma^2_4}{2\gamma_{12}\gamma_3},
	\end{equation}
	from which the solution is inferred at
	\begin{align}
	\varphi_3 &= \pi \mp \arccos \frac{\gamma^2_{12} + \gamma^2_3 - \gamma^2_4}{2\gamma_{12}\gamma_3} + 2\pi z_3 - \delta, \\
	\varphi_4 &= \pi \pm \arccos  \frac{\gamma^2_{12} - \gamma^2_3 + \gamma^2_4}{2\gamma_{12}\gamma_4} + 2\pi z_4 - \delta.
	\end{align}
	Again, the solution exists only when $\gamma_{12}$, $\gamma_{3}$, and $\gamma_{4}$ meet the triangle inequality, i.e., when $\gamma_\text{max}^r=\text{max}(\gamma_{12},\gamma_{3},\gamma_{4})$ is smaller than or equal to $(\gamma_{12}+\gamma_{3}+\gamma_{4})/2$.
	This procedure can be repeated for any number of terminals, always starting by rotating the loop around the zero to place the ($n-1$)st vertice on the $\mathcal{R}$-axis, that is, setting the vector $\bm{\gamma}_{1(n-1)}$ to be real.
	
	\begin{figure*}[ht]
		\includegraphics[width=1.0\columnwidth]{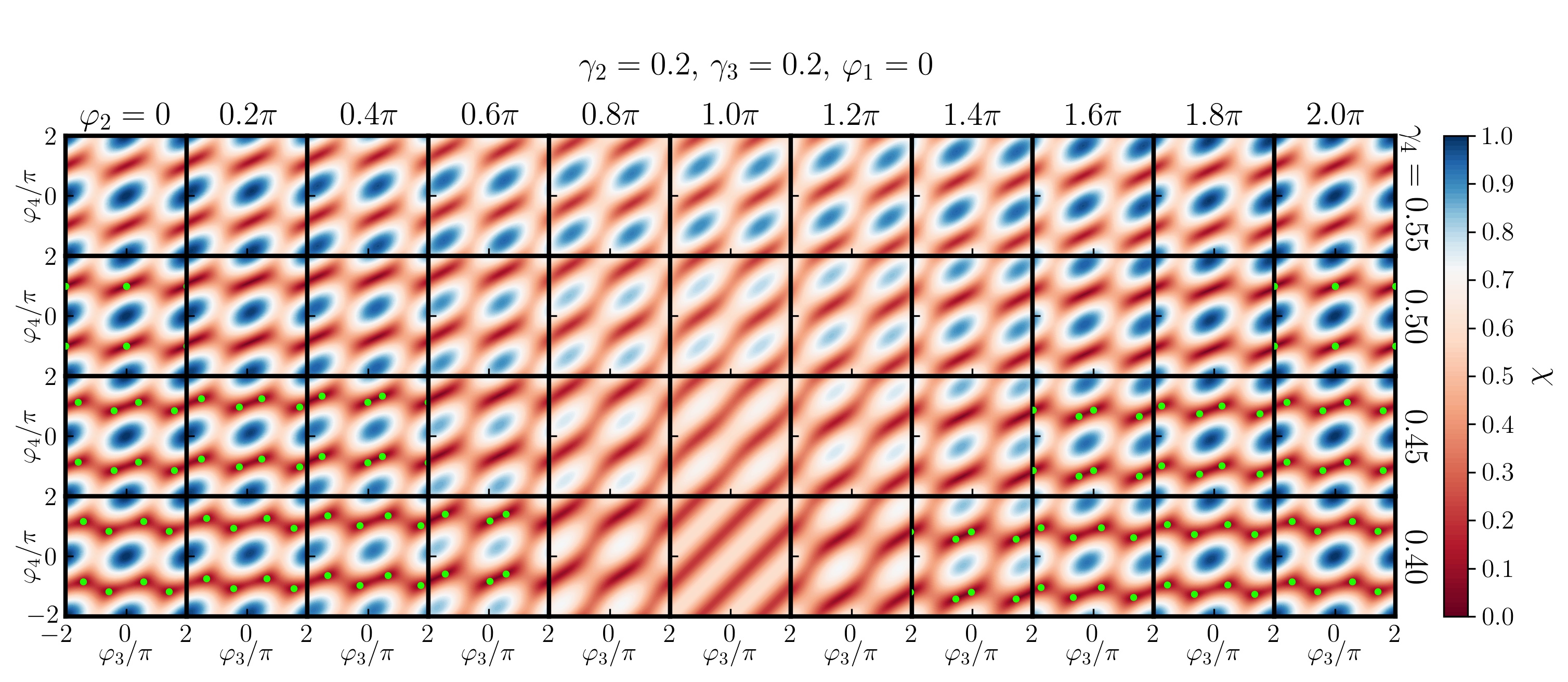}
		\caption{Example of the evolution of $\chi$ in a four-terminal setup in the $\varphi_3$-$\varphi_4$ plane for changing $\varphi_2$ and $\gamma_4$.  The rest of the parameters are set to $\gamma_2=0.2$, $\gamma_3=0.2$, $\varphi_1=0$. Green points show high-symmetry points $\chi=0$. The coloring is chosen in such a way that the maps can be read as phase diagrams for the case $\Gamma=\Delta$, $U=3\Delta$ for which $\chi^*=0.721$. Here, white represents the phase boundaries $\chi=\chi^*$, blue marks the singlet GS ($\chi>\chi^*$), and red a doublet GS ($\chi<\chi^*$).   
			\label{fig:chi4t}}
	\end{figure*}
	\begin{figure*}[ht]
		\includegraphics[width=1.0\columnwidth]{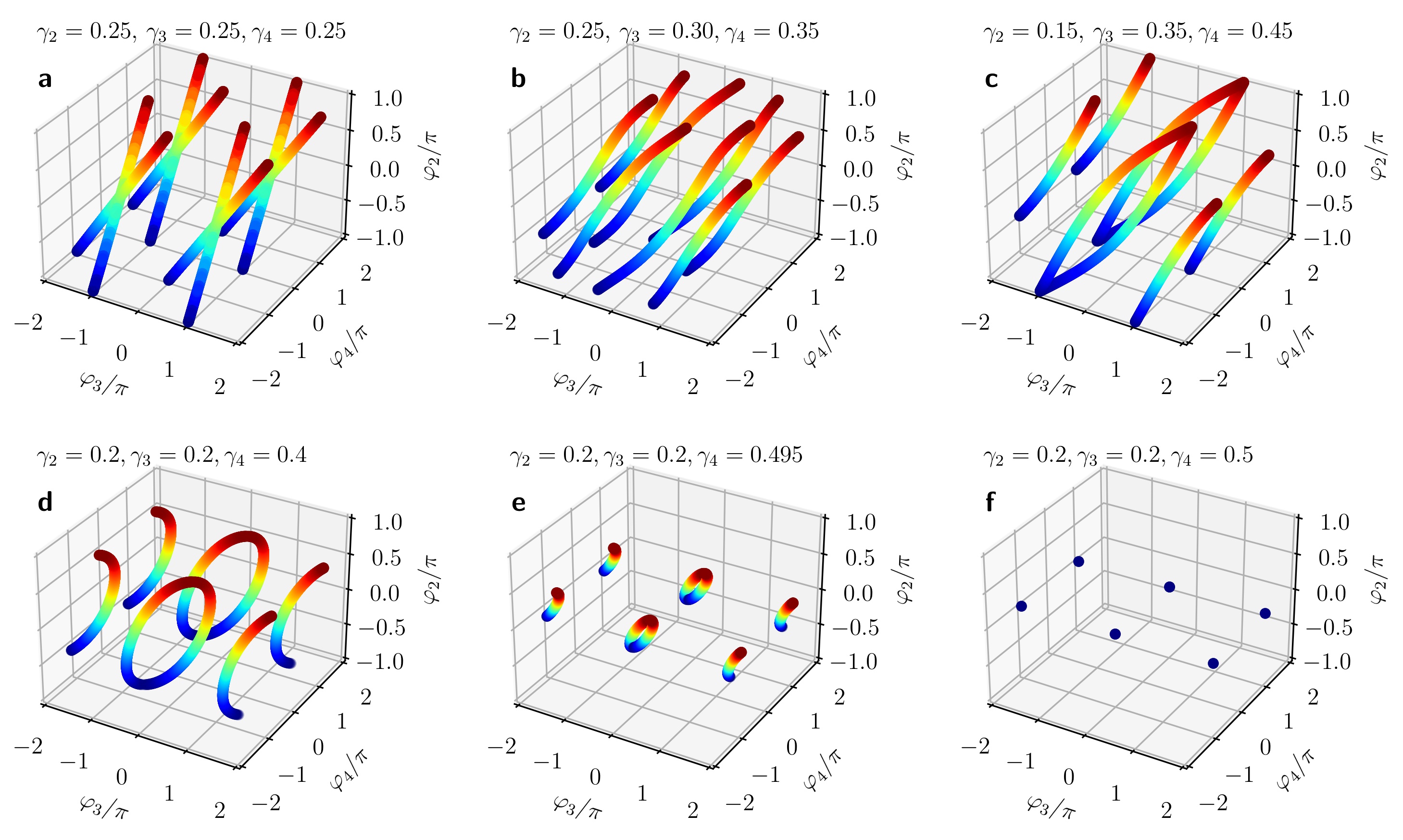}
		\caption{Examples of the $\chi=0$ solutions for a four-terminal setup in the $\varphi_3$ - $\varphi_3$ -$\varphi_2$ space for different combinations of $\gamma_2$, $\gamma_3$, $\gamma_4$ and with $\varphi_1=0$ in all panels. Note that a cube $-\pi \leq\varphi_j < \pi$ covers the first synthetic $3$-dimensional Brillouin zone. However, to better see the structure of the $\chi=0$ curves, we plot a broader range $-\pi \geq\varphi_{3,4} < \pi$. Also, the change in color of the curves signals the value of $\varphi_2$ for the sake of clarity. In the second row, the loops shrink as $\gamma_4$ approaches one-half, where there is only one isolated node in a Brillouin zone. There are no solutions above $\gamma_4>1/2$.
			\label{fig:chi0_3d}}
	\end{figure*}
	
	In the end, the same rigorous analysis achieved in the main text for three-terminal setups can be repeated for their four-terminal counterparts. However, the parameter space of four-terminal systems is much larger. In Fig.~\ref{fig:chi4t}, we thus present the evolution of $\chi$ for fixed $\gamma_2=0.2$, $\gamma_3=0.2$ ($\varphi_0\equiv0$), while varying $\gamma_4$ and $\varphi_2$. The coloring of the maps is chosen in such a way that it can be read as phase diagrams for the case $\Gamma=\Delta$, $U=3\Delta$ for which $\chi^*=0.721$. In particular, white signals phase boundaries, blue singlet ground state, and red the ground state. We will briefly return to the phases in the next section. Here, we focus on the positions of $\chi=0$ points as marked by green dots. First, note that the Brillouin zone is three dimensional with $-\pi\leq\varphi_2<\pi$,  $-\pi\leq\varphi_3<\pi$ and $-\pi\leq\varphi_4<\pi$ ($\varphi_1 \equiv 0$), but in Fig.~\ref{fig:chi4t} its cuts at fixed $\varphi_2$ are presented. Consequently, each cut contains two solutions for $\gamma_\text{max}^r<(\gamma_{12}+\gamma_{3}+\gamma_4)/2$, one for $\gamma_\text{max}^r=(\gamma_{12}+\gamma_{3}+\gamma_4)/2$, but none otherwise. While in the cuts, $\chi=0$ points are isolated they lie on continuous lines or closed loops, when the whole three-dimensional synthetic Brillouin zone is depicted as illustrated in Fig.~\ref{fig:chi0_3d} for various parameters. In the particular case of $\gamma_2=0.2$ and $\gamma_3=0.2$ shown in Fig.~\ref{fig:chi4t} and Fig.~\ref{fig:chi0_3d} the loops shrink as $\gamma_4$ approaches $1/2$. At such a special point, one of the relative couplings is exactly $1/2$, and we get a single finite energy node in the first Brillouin zone exactly at $\varphi_2 =0$, $\varphi_3=0$, and $\varphi_4 = \pm\pi$.  This can be straightforwardly generalized to any number of terminals. For $n>1$ the isolated node only exists in the case $\gamma_\text{max}\	\equiv\gamma_n=1/2$ and is placed at $\varphi_{j\neq n} = 0$ and $\varphi_{n} = \pm\pi$.

	\section{Supercurrent and universal current function $\mathcal{J}(\chi)$}
	The current operator $\hat{J}_j$ for a supercurrent flowing between the lead $j$ and the dot is defined as the time derivative of the particle number operator $\hat{N}_j$
	\begin{equation}
	\hat{J}_j = \partial_t \hat{N}_j= i\frac{e}{\hbar}\left[\hat{H},\hat{N}_j\right].
	\end{equation}
	Here, it is convenient to introduce a unitary transformation~\cite{Meden-2019}
	\begin{equation}
	c^\dagger_{j,k,\sigma}\rightarrow e^{-i\varphi_j/2}c^\dagger_{j,k,\sigma},\,\, c_{j,k,\sigma}\rightarrow e^{i\varphi_j/2}c_{j,k,\sigma},\
	\label{eq:trans}
	\end{equation}
	because it allow us to derive an alternative expressions to calculate the currents. In this way, the phases of the leads are moved to the hybridization term and can be included in the hoppings between the leads and the dot as
	\begin{equation}
	V^*_{\mathbf{k}j}\rightarrow e^{-i\varphi_j/2}V^*_{\mathbf{k}j},\,\, V_{\mathbf{k}j}\rightarrow e^{i\varphi_j/2}V_{\mathbf{k}j}.\
	\label{eq:tranhyb}
	\end{equation}
	Then, in equilibrium, the expectation value of the commutator of $\hat{N}_j$ with the (transformed) Hamiltonian term $H_{j,\text{SC}}$, which describes the superconducting lead $j$, vanishes due to the self-consistent definition of the BCS order parameter $\Delta =\sum_k \left< c_{j,k,\uparrow} c_{j,-k,\downarrow}\right>$. Therefore, the only relevant part of the commutator is $i\frac{e}{\hbar}\left[\hat{H}_{j,T},\hat{N}_j\right]$, which leads to the expression
	\begin{equation}
	\hat{J}_j=i\frac{e}{\hbar}\left[e^{i\varphi_j/2}\sum_{\mathbf{k},\sigma}V_{\mathbf{k},j}d^\dagger_\sigma c_{j,\mathbf{k},\sigma}-
	e^{-i\varphi_j/2}\sum_{\mathbf{k},\sigma}V_{\mathbf{k},j}^* c^\dagger_{j,\mathbf{k},\sigma}d_\sigma\right].
	\label{eq:Jdef_com}
	\end{equation}
	The same current operator can be obtained by the derivative of the transformed Hamiltonian $\hat{J}_j=\frac{2e}{\hbar}\partial_{\varphi_j} \hat{H}$. Now we can utilize the Hellmann-Feynman theorem. At zero temperature and with the chemical potentials of the leads set to zero, this gives an expression for the current between the lead $j$ and the dot as
	\begin{equation}
	J_j=\frac{2e}{\hbar}\frac{\partial E}{\partial \varphi_j},
	\label{eq:Jdef_prel}
	\end{equation}
	where $E$ is the ground state energy. Note that the reciprocal use of the Hellmann-Feynman theorem, that is, starting from Eq.~\eqref{eq:Jdef} and deriving the needed current operator, can be utilized in techniques such as numerical renormalization group (NRG) that work with transformed Hamiltonians (e.g. linear chains)~\cite{Saldana-2018}.
	
	We can now utilize the geometric factor $\chi$ to show that all currents in a general multiterminal setup follow from the equivalent two-terminal setup with the same $U$, $\Gamma$, $\chi$ and symmetric coupling. Because $E$ does not depend on the geometrical details of the leads, we can write	
	\begin{equation}
	J_j=\frac{2e}{\hbar}\frac{\partial E}{\partial \chi}\frac{\partial \chi}{\partial \varphi_j}=\mathcal{J(\chi)} \frac{\partial \chi}{\partial \varphi_j},
	\label{eq:Jdef}
	\end{equation}	
	where we introduced a universal factor $\mathcal{J}(\chi)$. It is a function of $\chi$ and depends parametrically on $U$ and $\Gamma$. The factor $\mathcal{J}(\chi)$ can be calculated from the ground state energy of the two-terminal symmetric setup. 
	
	In some techniques, it might be more convenient to directly calculate the current of the two-terminal setup instead of the ground-state energy. Because  current of the symmetric two-terminal setup can be calculated as $J_{n=2}=\mathcal{J}(\chi)\frac{\partial \chi_{n=2}}{\partial \varphi}$ with $\varphi=\varphi_2-\varphi_1$ we can use it to obtain the universal factor 
	\begin{equation}
	\mathcal{J}(\chi)=-\frac{2}{\sqrt{1-\chi^2}}J_{n=2}(\chi).
	\label{eq:universalJ}
	\end{equation}
	
\begin{figure*}[ht]
		\includegraphics[width=1.0\columnwidth]{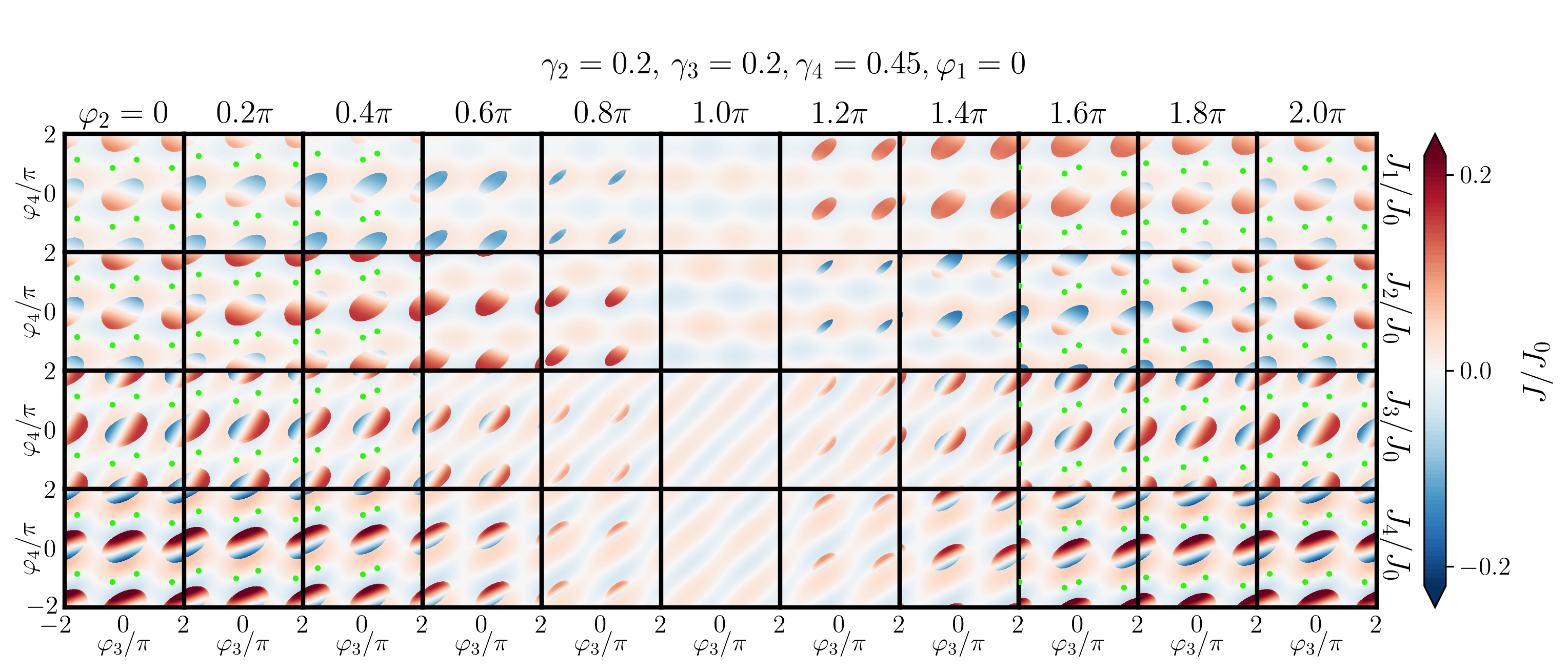}
		\caption{Example of the evolution of currents $J_1\dots J_4$ in a four-terminal setup in the $\varphi_3$-$\varphi_4$ plane for changing $\varphi_2$ where $J_0 = 2\Delta e /\hbar$.  Rest of the parameters are set to $\gamma_2=0.2$, $\gamma_3=0.2$, $\gamma_3=0.4$,  $\varphi_1=0$, $\Gamma=\Delta$ and $U=3\Delta$. Green points mark the position of the high-symmetry points $\chi=0$. 
			\label{fig:current4t}}
	\end{figure*}

	The main text deals in detail with the current in three-terminal setup. Here, we show some examples for the four-terminal system, as their full analysis is beyond our scope. Fig.~\ref{fig:current4t} presents maps of the four $J_j$ currents in the $\varphi_3$ and $\varphi_4$ planes for $\gamma_2=0.2$, $\gamma_2=0.2$, $\gamma_4 = 0.45$ and varying $\varphi_2$. The sharp edges between dark red (blue) and light blue (red) colors signal the position of the quantum phase transition between the doublet and singlet phases. This can be further confirmed by comparing the results for current with the corresponding phase diagrams in the third row of Fig.~\ref{fig:chi4t}. Obviously, there are parameter ranges, where the singlet phase forms only small isolated islands. Interestingly, within such islands, the currents from particular leads can have a different character. For example, the case with $\varphi_2=0.6\pi$ shows within the singlet island only positive (negative) current $J_{1}$ ($J_{2}$) bud $J_{3}$ ($J_{4}$) alternates sigh. This reflects the changes in the geometric factor $\partial_{\varphi_j} \chi$. Nevertheless, the effect requires a thorough analysis, which is, however, beyond the scope of the present study.
	
	\section{NRG implementations \label{sec_app_nrg} }

	A standard two-channel NRG approach to superconducting problems has been implemented within the open source code of NRG Ljubljana with $z$-averaging employed at $N_z=4$ \cite{NRGzenodo}. A maximum of $2000$ states have been kept during diagonalization, and we set $\Lambda=4$. 
	
\begin{figure*}[ht]
		\includegraphics[width=1.0\columnwidth]{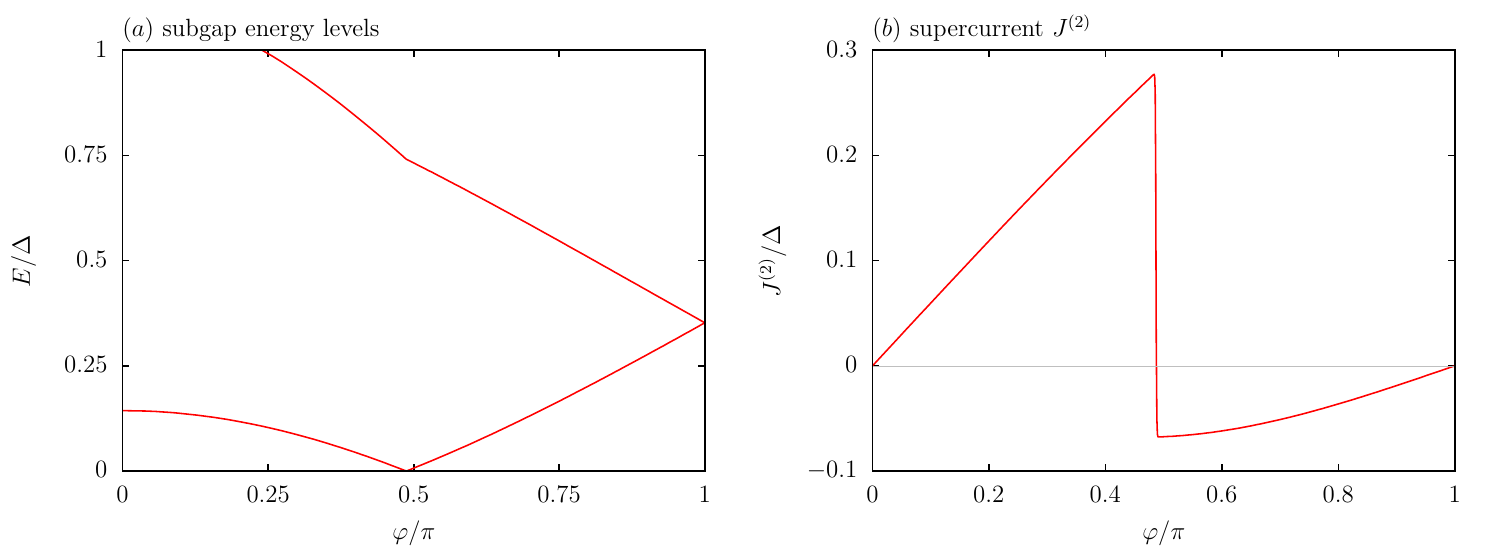}
		\caption{
			(a) 
			Evolution of subgap energy levels with phase bias $\varphi$ of a symmetric SC-AIM with $U=3\Delta$, $\Gamma=\Delta$ and $\Delta=0.0005B$. At $\varphi \approx0.487\pi$ there is a crossing of the ground state with the first excited state. Here a QPT takes place.
			(b)
			Phase dependence of the current $J^{(2)}$ for the same parameter as in panel (a). The QPT point is visible as a jump from positive to negative values of the current.
			\label{fig:nrg}}
	\end{figure*}

	The correspondence between $\chi$ and the two-channel phase bias is established on the basis of \eqref{eq:two_channel} as $\chi = \cos (\varphi/2)$. Using $U=3\Delta, \Gamma=\Delta$ and $\Delta=0.0005B$ with bandwidth $2B$, we obtained the corresponding NRG spectrum and additionally measured the Josephson current. The resulting subgap energy levels and current $J^{(2)}$ are shown in Fig.~\ref{fig:nrg}(a) and (b), respectively. From the former, a critical phase of $\varphi_c \approx 0.487\pi$ was determined, which gives the desired value of critical $\chi_c=0.721$. The supercurrent $J^{(2)}$ then serves as an input to determine the universal function $\mathcal{J}(\chi)$ via Eq.~\eqref{eq:Jdef}.

%